\documentclass{jaa}

\usepackage[utf8x]{inputenc}
\usepackage{ upgreek }
\usepackage{fullpage}
\usepackage{multicol}
\usepackage{multirow}
\usepackage{graphicx}
\usepackage{verbatim}
\usepackage{epstopdf}

\def\be{\begin{equation*}}
\def\ee{\end{equation*}}
\def\ba{\begin{eqnarray}}
\def\ea{\end{eqnarray}}

\begin{document}

%%paper title
%%For line breaks \\ can be used within title 
\title{The dynamic spectral signatures from Lunar Occultation: A simulation study\\}

%%author names are separated by comma (,) 
%%use \and before the last author name 
%%use a * along with the number separated by comma
%% for the  author for correspondence
%%\textsuperscript{number} is used for affiliation
%%\affilOne, \affilTwo etc., upto \affilTwentyfive is possible
%%Please note the first letter after \affil is capitalised in the command
%%

\author{Jigisha V. Patel\textsuperscript{1}, Avinash A. Deshpande\textsuperscript{1}}
\affilOne{\textsuperscript{1}Raman Research Institute, C. V. Raman Avenue, Sadashivanagar, Bangalore - 560 080, India.\\}

%%escape two column mode for title, affiliation and abstract
%%by giving \twocolumn command as shown

\twocolumn[{

\maketitle

%%include \corres to print the corresponding author Email id
\corres{jigishapatel.7793@gmail.com}    %, desh@rri.res.in}

%%include \msinfo for
%%manuscript information such as
%%received, revised and accepted dates
%%

%%abstract
\begin{abstract}
Lunar occultation, which occurs when the Moon crosses sight-lines 
to distant sources, has been studied extensively through apparent 
intensity pattern resulting from Fresnel diffraction, 
%which depends on the angular size of the source, 
%the frequency of observation and distance to the obstruction. 
%Such lunar occultation observations have been successfully 
 and has been successfully
used to measure angular sizes of extragalactic sources. 
However, such observations to-date have been mainly over 
narrow bandwidth, or averaged over the observing band, 
and the associated intensity pattern in time has rarely been 
examined in detail as a function of frequency over a wide band. 
Here, we revisit the phenomenon of lunar occultation with 
a view to study the associated intensity pattern as a function 
of both time and frequency. Through analytical and simulation approach, 
we examine the variation of intensity across the dynamic spectra, 
and look for chromatic signatures which could appear as discrete 
dispersed signal tracks, when the diffraction pattern is adequately 
smoothed by a finite source size. We particularly explore circumstances 
in which such diffraction pattern might closely follow the interstellar 
dispersion law followed by pulsars and transients, such as the 
Fast Radio Bursts (FRBs), which remain a mystery even after a decade 
of their discovery. In this paper, we describe details of this 
investigation, relevant to radio frequencies at which FRBs have 
been detected, and discuss our findings, along with their implications.
 We also show how a {\it band-averaged} light curve suffers from
temporal smearing, and consequent reduction in contrast of intensity
variation, with increasing bandwidth. We suggest a way to recover 
the underlying diffraction signature, as well as the sensitivity 
improvement commensurate with usage of large bandwidths.

\end{abstract}

%%insert keywords separated by 3 hyphens using \keywords{words}
% -- techniques: spectroscopic

}]
%%close the twocolumn escape here

%%include \doinum{number}for the DOI number in the header
%%include \volnum{number} for the volume number in the header
%%include \year{yyyy} for  year of publication in the header
%%include \pgrange{num--num} page range of article in the header
%%include \artcitid{num} for the article citation id
%%include \lp to print last page of the article
%%include \setcounter{page}{pagenum} for the exact starting page of the article

\doinum{12.3456/s78910-011-012-3}
\artcitid{\#\#\#\#}
\volnum{123}
\year{2016}
\pgrange{1--5}
\setcounter{page}{1}
\lp{25}

% 0.5 second per arcsecond is the Moon motion (relative to the astro source)
% sqrt(2*lambda/D) for 1m wavelength is 14.4 arcsec
% sqrt(2*lambda/D) for 20 cm wavelength is 6.5 arcsec
\section{Introduction}

An occultation takes place when a nearby celestial object 
moves in front of a distant one and blocks it from view. 
In early years, these observations in radio frequencies yielded
measurements of diameters of small radio sources with 
accuracies much better than any other method available then, while 
requiring less instrumental efforts ( Scheuer 1962, Von Hoerner 1963). 
The fluctuations in the received power
when the Moon obstructs a radio source depends on the size of the source
relative to the size of the Fresnel zone at the Moon's distance, which over the
range of frequencies of interest, ranges from a few arc-second to a maximum
of about 20 arc-second (Hazard et al. 1963,1976; Scheuer 1962).    
%; Singal 1987). 
 Lunar occultation observations for about 1000 radio
%more than 150 radio 
sources have been made with the steerable Ooty Radio Telescope at 
Ootacamund, India (Swarup 1971; see also, Singal 1987 and references therein) 
However, all the findings 
so far used band averaged signatures of occultation light curves.

Here, we have attempted to find the spectral imprints of 
Lunar occultation, which over a wide range of frequencies 
has not been explicitly studied yet. The intensity pattern 
due to Fresnel diffraction by straight edge for a range of 
frequencies depict spatial dispersion of fringes. 
 This pattern sweeping across an observer,
due to the motion of the moon relative to the sight-line to the
source, manifests into temporal dispersion.
Hence, the observed intensity profiles
will show relative time delay that is correlated with frequency,
when corresponding features at different radio frequencies are compared.
The apparent pattern 
can be calculated using the standard formula for 
intensity at any point due to Fresnel diffraction by straight edge, 
resulting into dispersed dynamic spectrum. We have attempted 
to find if the dispersion signature observed follows 
a particular trend and explored the possibility of changes 
in the trend under different circumstances. 
We have in particular explored the possibility of 
comparing it with the ISM dispersion law, followed by pulsars, 
FRBs (Lorimer et al. 2007) or similar pulsed signals.   

In section 2, we have first deduced a generic expression 
of intensity as function of frequency and distance, which is 
further utilized to find dispersion like delay and for simulating 
the spectral signatures. Further, we take into account few 
situations and observe the changes in the diffraction pattern, 
in turn the changes in dynamic spectrum and dispersion characteristics. 

In section 3, situations differing in assumptions of 
velocity of obstruction, source size and minimum distance 
to obstruction are considered. The dispersion delay and 
spectral signatures for above scenarios are analyzed using 
simulations. We have also tried to explore circumstances 
in which the dispersion trend found for diffraction closely 
follows the dispersion trend of ISM.

 In section 4, we discuss some key implications of wide bandwidth
observations for Signal-to-Noise ratio improvement in
estimation of the average intensity pattern  
and summarize our conclusions in the last section.

\section{Spatial dispersion in Fresnel diffraction}

An interference or diffraction pattern is created when 
an object partially blocks the path from a monochromatic light source. 
Constructive interference occurs when there is a path length 
difference which is an integral multiple of the wavelength 
of the light from the source. When the obstructing object is 
a straight edge, the interference
pattern will occur on one side only and caused by a path 
difference between light coming from a point source.  

Lunar occultation is one such example of frequent and naturally 
occurring Fresnel diffraction by straight edge (Ghatak 2010). 
Since the astronomical sources are very far, they appear point-like
and a small part of lunar limb acts as straight edge for the distant source. 
The intensity $I$ at any distance $x$ from geometric shadow 
for point source can be given by\footnote{This particular form of 
expression is similar to that given by Richmond (2005; ``Diffraction 
effects during a lunar occultation"; 
http://spiff.rit.edu/richmond/occult/bessel/bessel.html)}, 

\begin{multline}
I=\int_{-\infty}^xexp\left(\frac{i\pi u^2}{L\lambda}\right)du \int_{-\infty}^xexp\left(\frac{-i\pi u^2}{L\lambda}\right)du  \\ 
\end{multline}

where, $\lambda$ is the wavelength of source light, 
$x$ is the distance away from the edge, and 
$L$ is the distance between the sight-line and the obstruction. 
In case of lunar occultation, $L$ is taken as 384,000 km.

Above intensity equation for a point-like source is in 
integral form, and can be further simplified to derive relation 
of intensity as function of time and distance. The field $E$ at 
a point in observer plane, corresponding to $x$, can be given by,
\begin{equation}
E \ = \int_{-\infty}^xexp\left(\frac{i\pi u^2}{L\lambda}\right)du
\end{equation}

%\begin{equation}
% \mspace{20mu}  = \int_{-\infty}^0exp\left(\frac{i\pi u^2}{L\lambda}%\right)du + \int_0^xexp\left(\frac{i\pi u^2}{L\lambda}\right)du
%\end{equation}

%where,
%\begin{equation}
% u=\sqrt{\frac{L\lambda}{\pi}}exp\left(\frac{i\pi}{4}\right)y
%\end{equation}
%On substituting the expression for variable u,
%\begin{multline}
%E = \sqrt{\frac{L\lambda}{\pi}}exp\left(\frac{i\pi}{4}\right)\int_{-\infty}%^0exp(-y^2)dy+ \\
 %\sqrt{\frac{L\lambda}{\pi}}exp\left(\frac{i\pi}{4}\right)\int_{0}^z exp(-y^2)dy
%\end{multline}
%where $ z= \sqrt{\frac{\pi}{L\lambda}}exp\left(\frac{-i\pi}{4}\right)x$ 
%On simplifying,
%\begin{equation}
%E = \sqrt{\frac{L\lambda}{\pi}}exp\left(\frac{i\pi}{4}\right)\frac{\sqrt\pi}{2}+ \\
%\sqrt{\frac{L\lambda}{\pi}}exp\left(\frac{i\pi}{4}\right)\frac{\sqrt\pi}{2}%erf(z)
%\end{equation}
%Normalizing the $E$ by its value $E_0$ without obstruction,
%\begin{equation}
%\frac{E}{E_0}= \frac{(1+erf(z))}{2}
%\end{equation}%

The normalized intensity at any point is equal to 
$\left|\frac{E}{E_0}\right|^2$, 
 where $E_0$ is the value of the field observed without obstruction.
When a broadband source is considered, the same intensity 
at two different wavelengths will appear at corresponding 
different distances $x_1$ and $x_2$ from the edge such 
that the field values $E_1$ and $E_2$ are results of similar 
interference. Here, x = 0 at all frequencies corresponds to the 
geometrical shadow, when the (point) source is on the Moon’s limb. 
Considering this, the relation between distance and frequency 
$f_1$ and $f_2$ can be found out. 

This would require $z_1$=$z_2$, implying
\begin{equation}
\frac{x_1}{x_2}=\sqrt{\frac{\lambda_1}{\lambda_2}}=\sqrt{\frac{f_2}{f_1}}
\end{equation}
The difference $\Updelta x = (x_2-x_1)$ between these distances 
(corresponding to same intensity feature, 
but at different frequencies) is thus,  
\begin{equation}
\Updelta x = {x_2}{\sqrt f_2}\left(\frac{1}{\sqrt f_2}-\frac{1}{\sqrt f_1}\right) 
\end{equation} 
From the above equation, it can be deduced that the spatial 
dispersion between two points on intensity curves is inversely 
proportional to square root of the frequency. 

\section{Temporal dispersion}

In the above formulation, we have implicitly considered 
the obstruction to be stationary. For an occulting obstruction 
moving in a particular fashion, the resulting dispersion pattern 
will translate to temporal dispersion. In the following subsections, 
the dynamic spectrum for different relative motions of obstruction 
for both, point source and finite width source, are discussed.   

\subsection{The velocity of obstruction is constant}

In order to get dispersion law in the fashion described usually 
for the ISM, the obstruction can be considered to be moving 
with uniform velocity. The new formula associating time and 
frequency can be derived simply by dividing Equation 9 by 
constant value of velocity.

For the case of Lunar occultation, the source 
(which is farther compared to the obstruction) 
is considered stationary. Hence, for the obstruction 
moving with velocity v m/s, 
\begin{equation}
\Updelta t = t_2 - t_1 = {t_2}{\sqrt f_2}\left(\frac{1}{\sqrt f_2}-\frac{1}{\sqrt f_1}\right) 
\end{equation} 
where $\Updelta t$ is the time delay between same intensity 
feature observed at two frequencies. $t_1$ and $t_2$ are the 
associated time instants (measured from the time occultation began) 
at frequencies $f_1$ and $f_2$, respectively. It can be concluded 
from above equation, that as the time increases the dispersion 
delay also increases. The Figure 1 is the dynamic spectrum simulated 
for lunar occultation for a point source with 1200 MHz to 1500 MHz 
frequency range. The dynamic spectrum shows that secondary fringes 
farther in time are more dispersed and less intense.   

\begin{figure}
\includegraphics[width=1.0\linewidth]{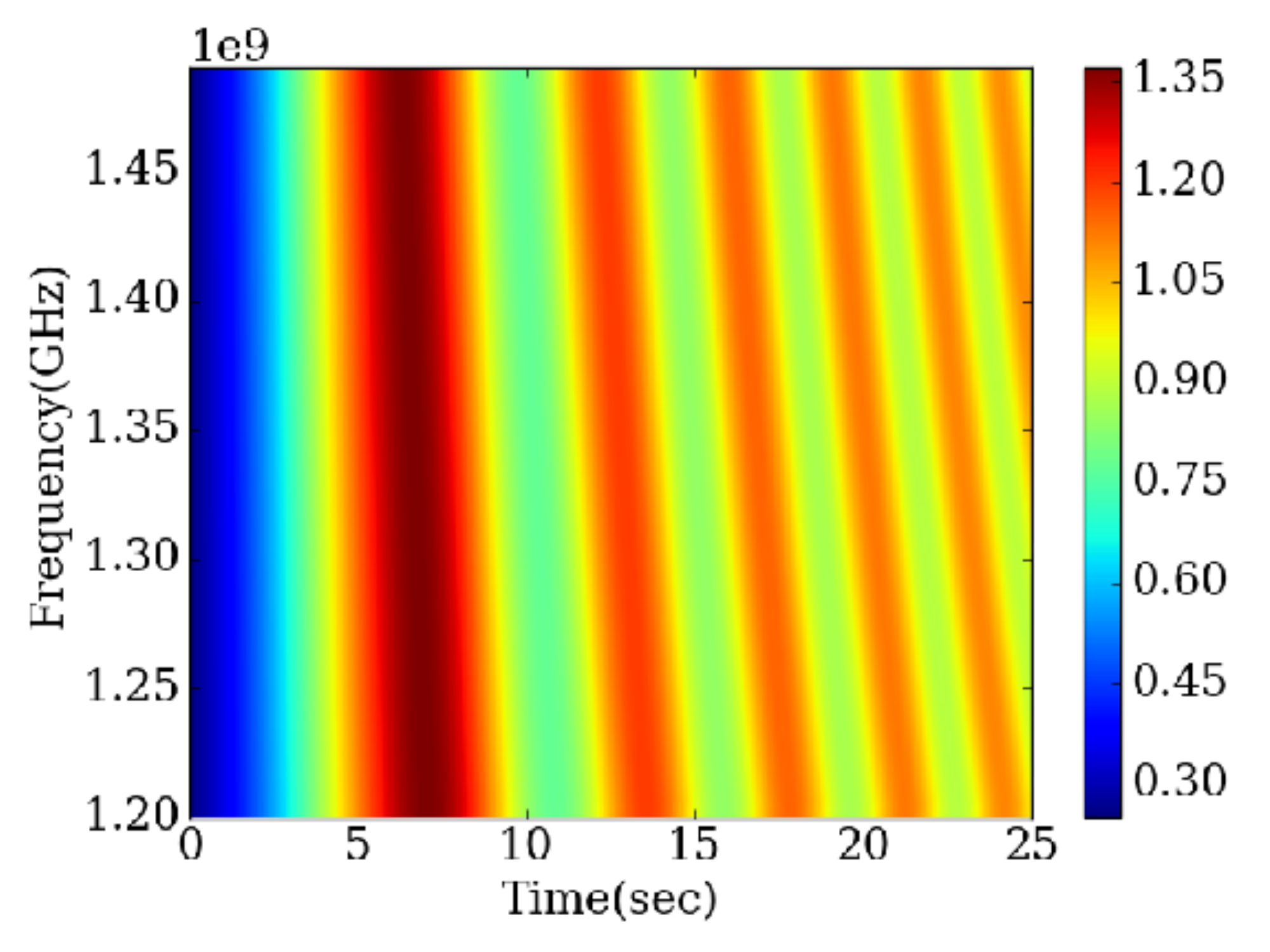}
\caption{Intensity pattern of lunar occultation for a point source 
for uniform motion of the Moon}
\label{fig:test}
\end{figure}

The motion of obstruction involves two cases, namely ingress 
and another is egress. These two cases follow different, 
in fact opposite dispersion trends. In
order to make the effects clearly evident a very broad band 
of 1200 MHz to 3000 MHz is selected. Diffraction patterns 
in Figure 2 are for the motion of the Moon relating both 
ingress and egress. Considering center of lunar disk as origin, 
the negative and positive time indicates the time before and 
after the source crosses from left to right side of origin, 
respectively. The ingress shows negative dispersion in which 
a given intensity peak at a frequency appears to arrive earlier 
than that at the  lower frequency. Far in the geometric shadow 
behind the lunar disk also there is finite intensity but of micro scale.

\begin{figure}[h]
\centering
\includegraphics[width=1.0\linewidth]{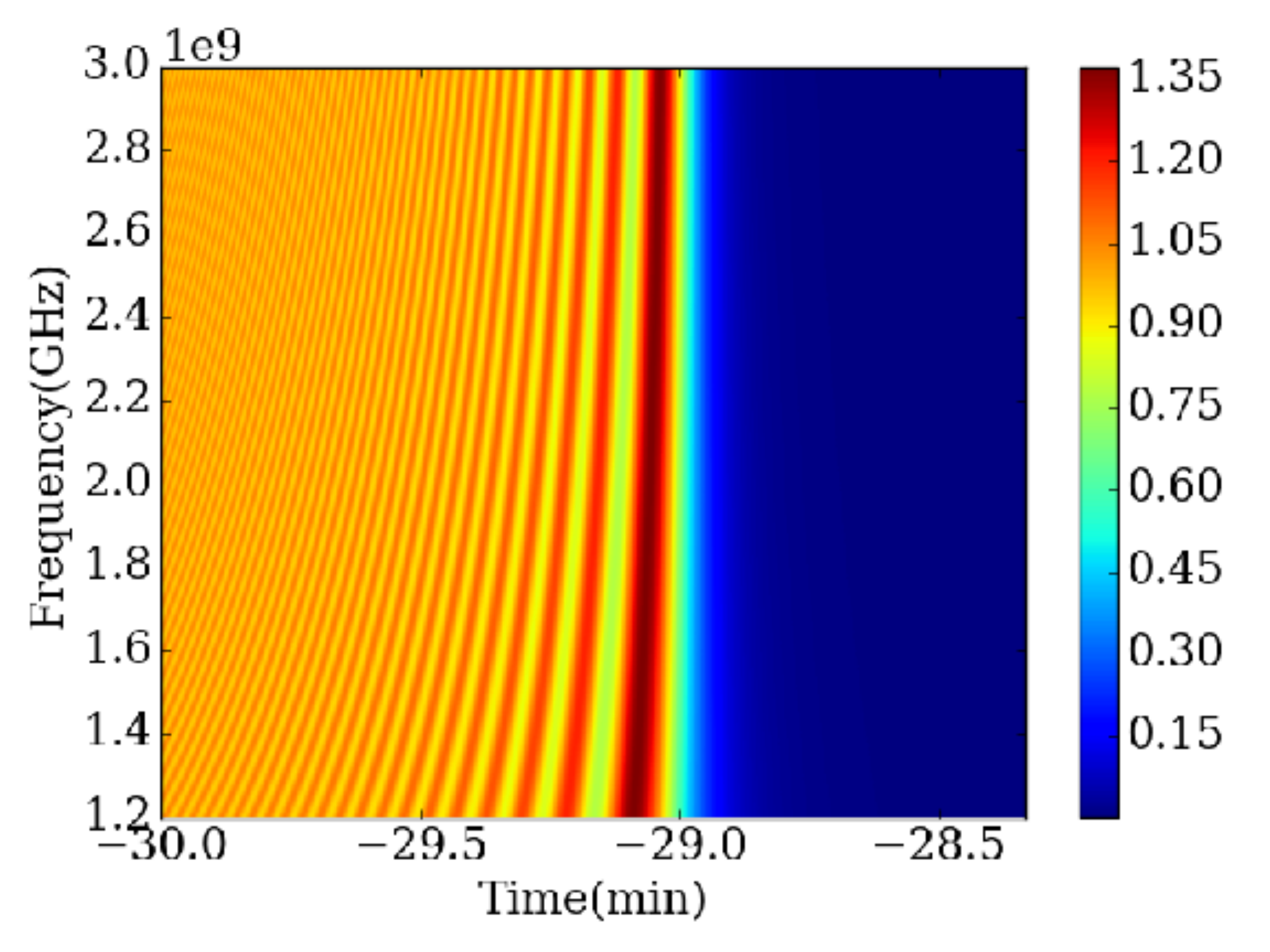}
\includegraphics[width=1.0\linewidth]{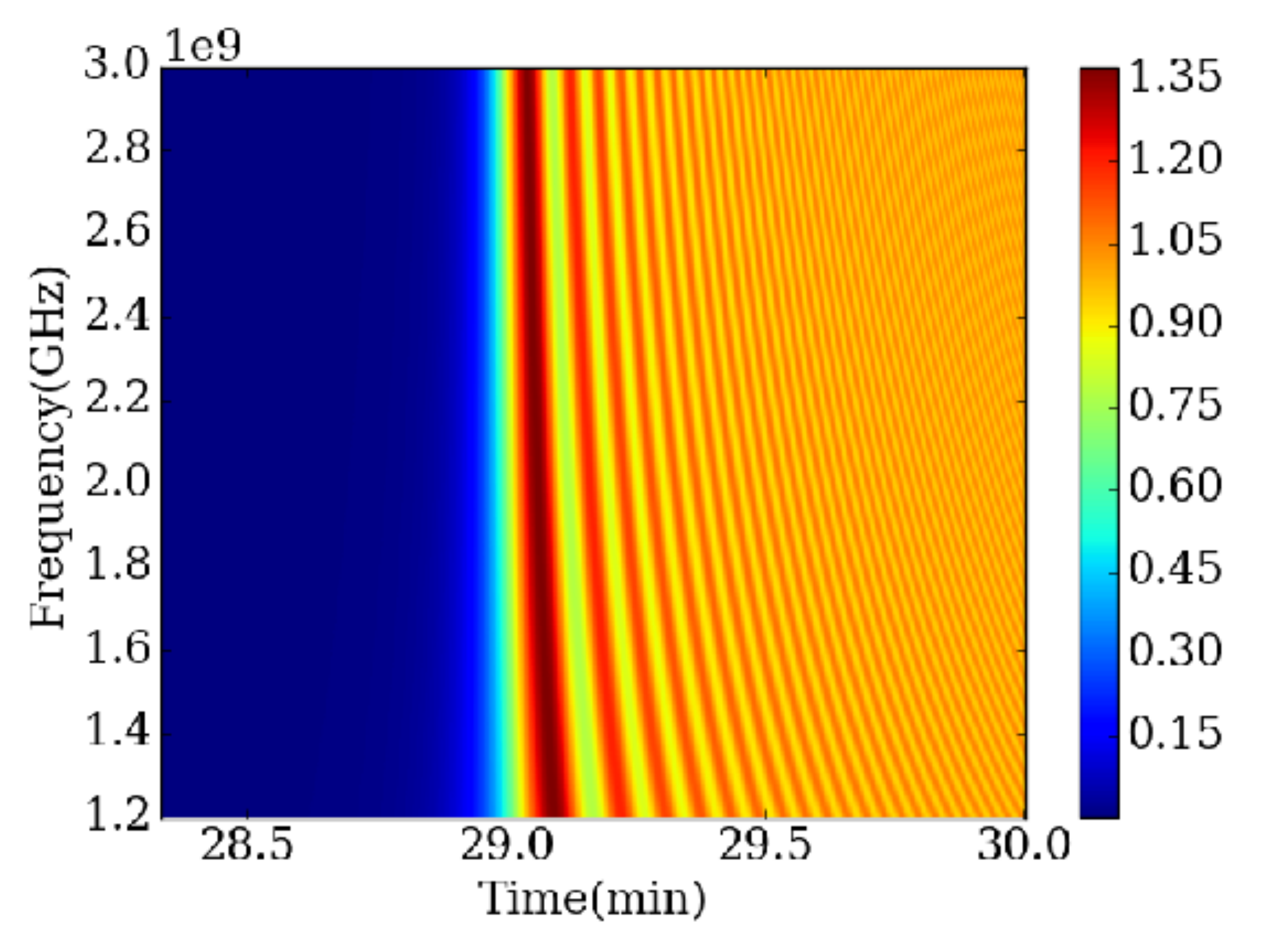}
\caption{The dynamic spectra pre and post shadowing, 
wherein {\it negative} dispersion during ingress (top panel) 
and {\it positive} dispersion during egress (bottom panel) are clearly apparent.}
\label{fig:test}
\end{figure}

\subsection{Obstruction moving with uniform deceleration}

In order to explore if the ISM-like dispersion trend 
(for ISM, the time delay at a given frequency is proportional 
to inverse square of frequency) can be mimicked by diffraction, 
nonlinear motion of obstruction is considered. 
Constant deceleration is a common phenomenon which might 
be experienced by near earth asteroids, possibly large sized 
satellites due to drag forces in low earth orbit, 
or the objects moving in highly elliptical orbit. 
If an obstruction, moving with constant deceleration 
is considered, then with the appropriate parameters the 
ISM dispersion law can be achieved.

Let us assume the following time evolution of x,
\begin{equation}
x=x_0+vt-\frac{1}{2}at^2
\end{equation} 
where v and a are velocity and deceleration respectively.
The roots of the above quadratic equation are,
\begin{equation}
t=\frac {v\pm\sqrt{v^2-2a(x-x_0)}}{a}
\end{equation} 
\begin{equation}
t =\frac{v}{a}\pm \frac{v\sqrt{1-(2a(x-x_0)/v^2)}}{a}
\end{equation} 

The second solution, after expanding the square root term 
using binomial expansion, and assuming \\ $ (x-x_0) =y$, 
may be expressed as
\begin{equation}
\mspace{20mu}=\frac{v}{a}\left(\frac{1}{2}-\sum_{n=1}^{\infty} {\frac{1}{2}\choose n} \left(\frac{-2ay}{v^2}\right)^n\right)
\end{equation}

In order to find the dispersion law followed by diffraction 
pattern under different situations, the following form 
is assumed with suitable units, so as to make it comparable 
with ISM dispersion law.
\begin{equation}
\Updelta t(ms) = K \\ * \\ K_{DM} \left({(f_1/(GHz))^{-\alpha}}-{(f_2/(GHz))^{-\alpha}}\right) 
\end{equation} 
Here, $K$ = 4.15 $(GHz)^\alpha$ ms, $\alpha$ is the power-law index, 
and
$K_{DM}$ is the {\it proportionality constant equivalent to 
ISM dispersion measure (DM)}.
 
%\vspace{1cm}

It is worth noting that, given the basic dependence 
$x \propto f^{−0.5}$,
if we were to obtain the ISM-like dispersion law 
$t \propto f^{-2}$, we necessarily require
$t \propto x^4$. 
We have explored a range of combinations of a,x,v and 
initial position in such a way that higher-order terms 
can be neglected, to see if we can get effectively a relation 
similar to that of ISM dispersion law. For example, if we take 
the distance to the obstruction as 6000 km, deceleration 
a= 17.4 $m/s​^2$ and initial velocity 200 m/s, ignoring 
if these are or not realistic, then in the expansion as 
in Equation 14, the terms for n$\ge$5 play insignificant role, 
and can therefore be neglected. In Figure 3, the time instants 
corresponding to the maximum intensity track sampled at different 
frequencies are plotted, assuming the above mentioned model parameters. 
In Figure 4, the top panel depicts the maximum amplitude of the 
band-summed feature (obtained after the otherwise dispersed {\it pulses} 
align maximally on {\it dedispersion}, following an assumed law), 
as a function of the trial $\alpha$, the power-law index of frequency.
The bottom panel of Figure 4 shows the corresponding optimal values
of the proportionality constant $K_{DM}$ as a function of $\alpha$.
%Figure 4 depicts the power-law index of frequency and proportionality 
%constant for which all the otherwise dispersed 'pulses' align maximally on dedispersion,
%and results into maximum amplitude of intensity summed over all frequencies. 
%The logarithmic peaks in the plot of amplitude vs. Proportionality constant 
%is due to unequal sampling of power-law index and proportionality constant.  

\begin{figure}
\centering
\includegraphics[width=1.0\linewidth]{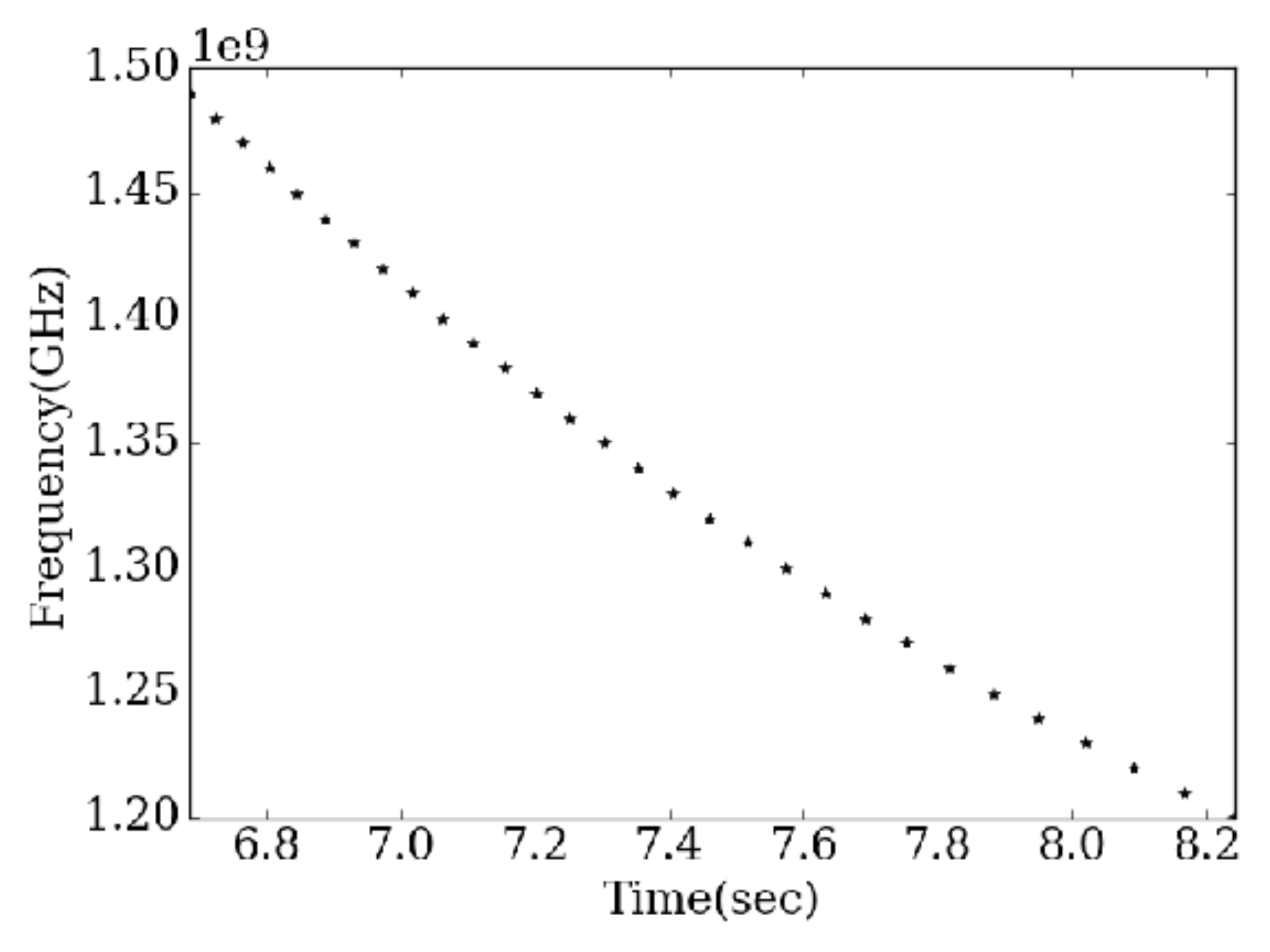}
\caption{Dispersion of peak intensity points of 
first Fresnel lobe when the  obstruction is moving with 
constant deceleration for 1200-1500 MHz band}
\label{fig:test}
\end{figure}

\begin{figure}
\centering
\includegraphics[width=1.0\linewidth]{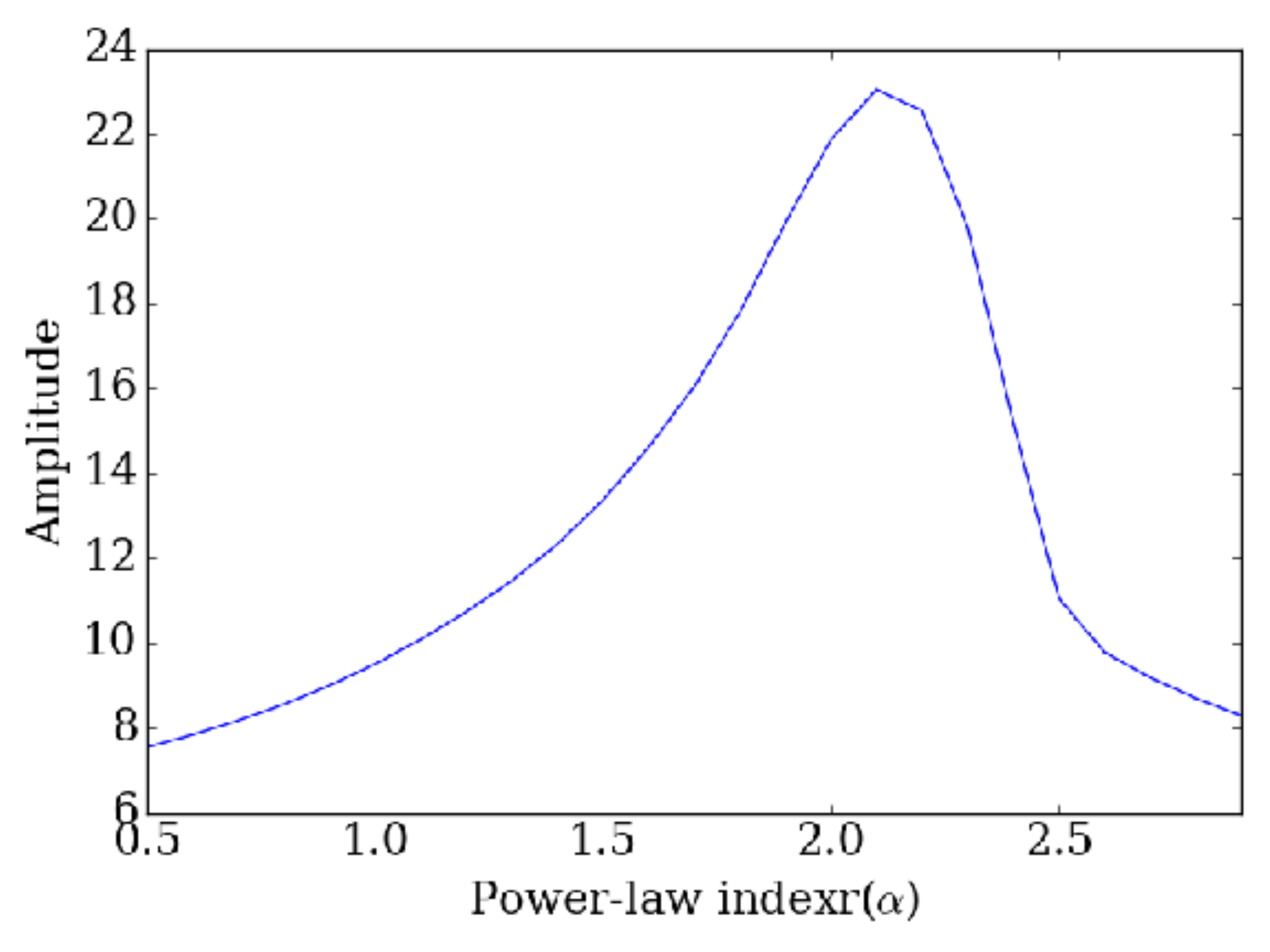}
\includegraphics[width=1.0\linewidth]{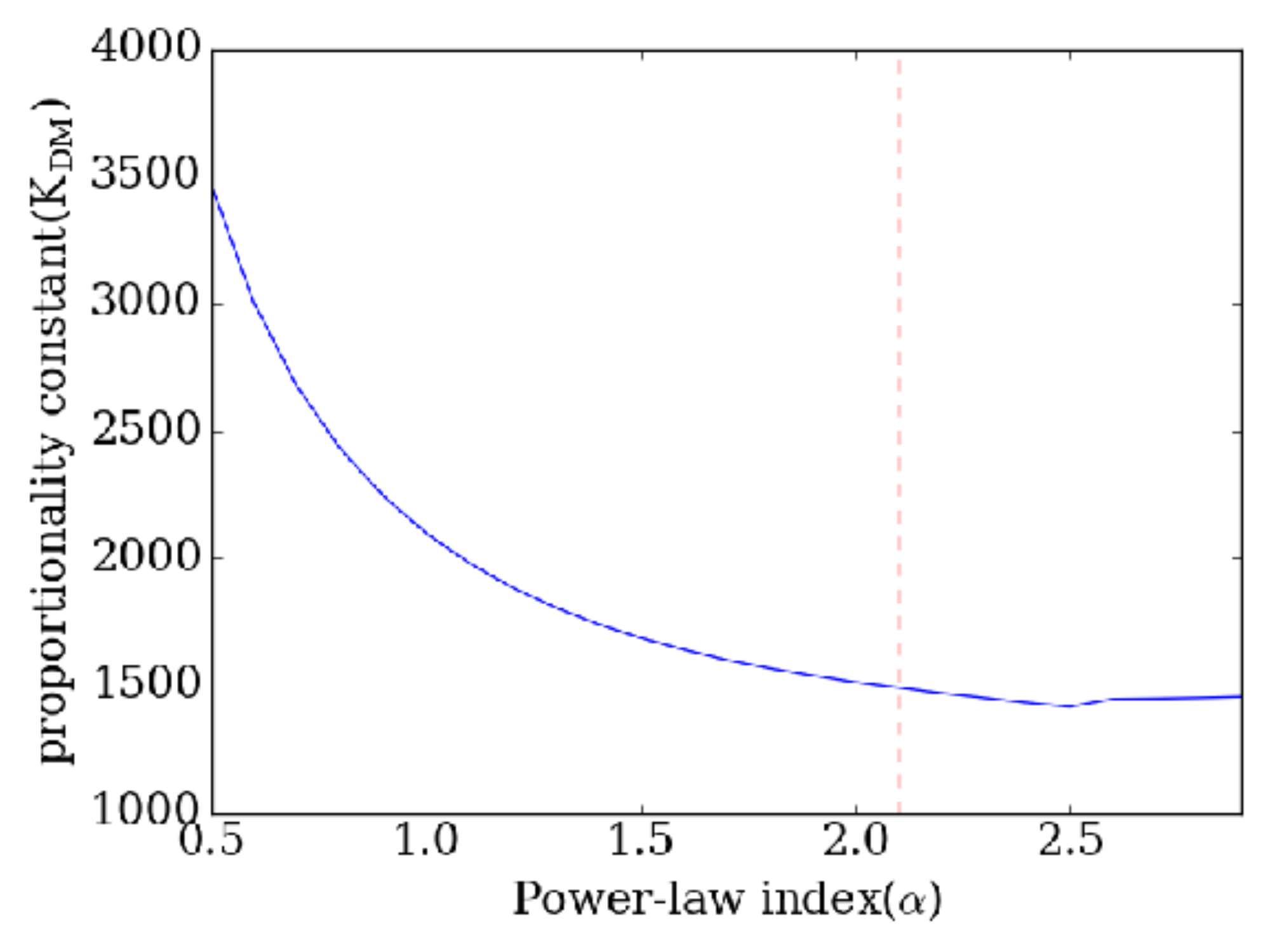}
\caption{Dependence of amplitude of (band-summed) intensity 
(top panel), and the corresponding optimal proportionality 
constant $K_{DM}$ (bottom panel), on the trial power-law index 
are shown. These refer to the case when effect of deceleration is included.}
\label{fig:test}
\end{figure}

\subsection{Grazing occultation}

\begin{figure}
\centering
\includegraphics[width=0.8\linewidth]{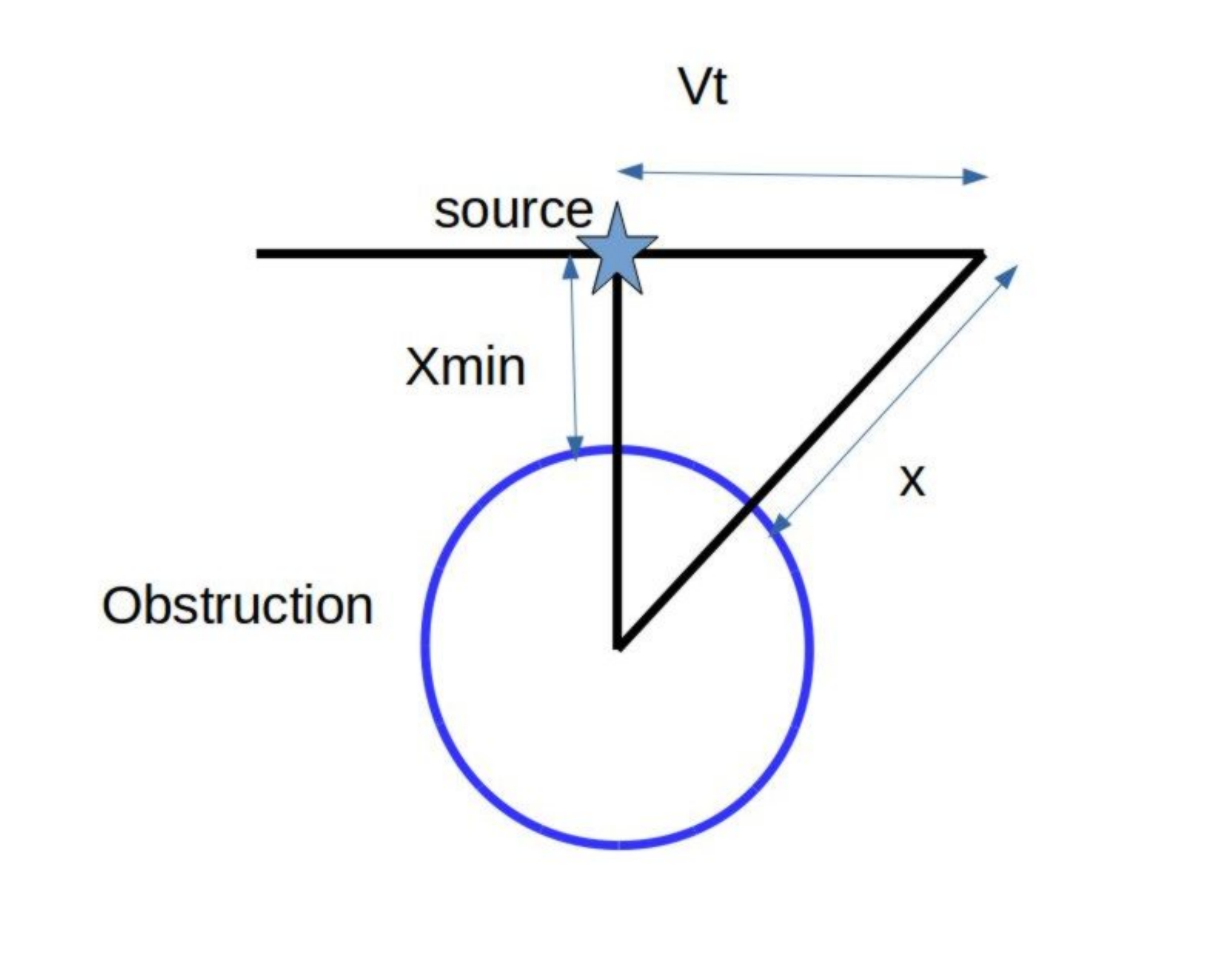}
\caption{Source at some distance from obstruction}
\label{fig:test}
\end{figure}

\begin{figure*}[h]
\begin{multicols}{2}
    \includegraphics[width=1.0\linewidth]{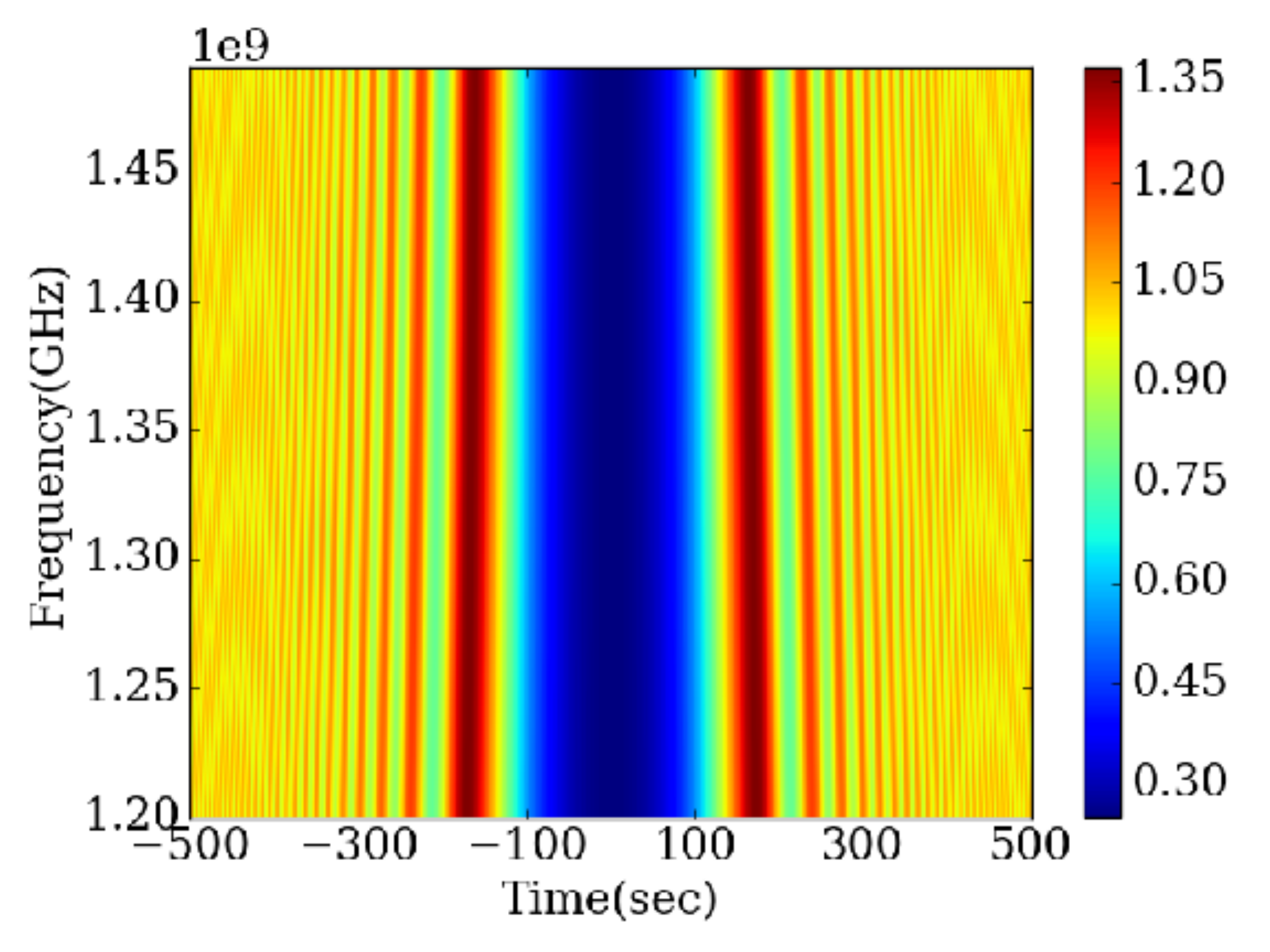}\par 
    \includegraphics[width=1.0\linewidth]{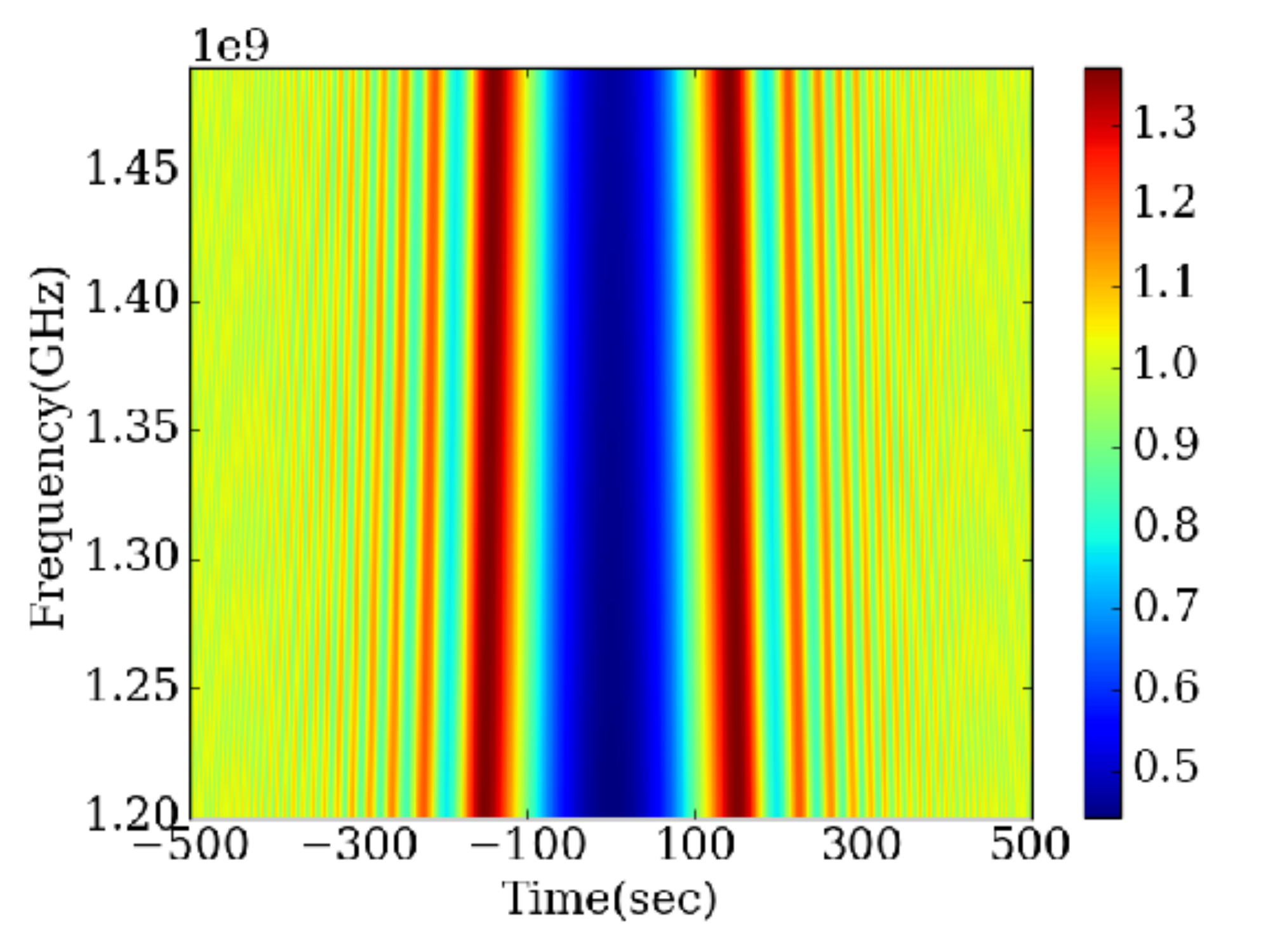}\par 
    \end{multicols}
\begin{multicols}{2}
    \includegraphics[width=1.0\linewidth]{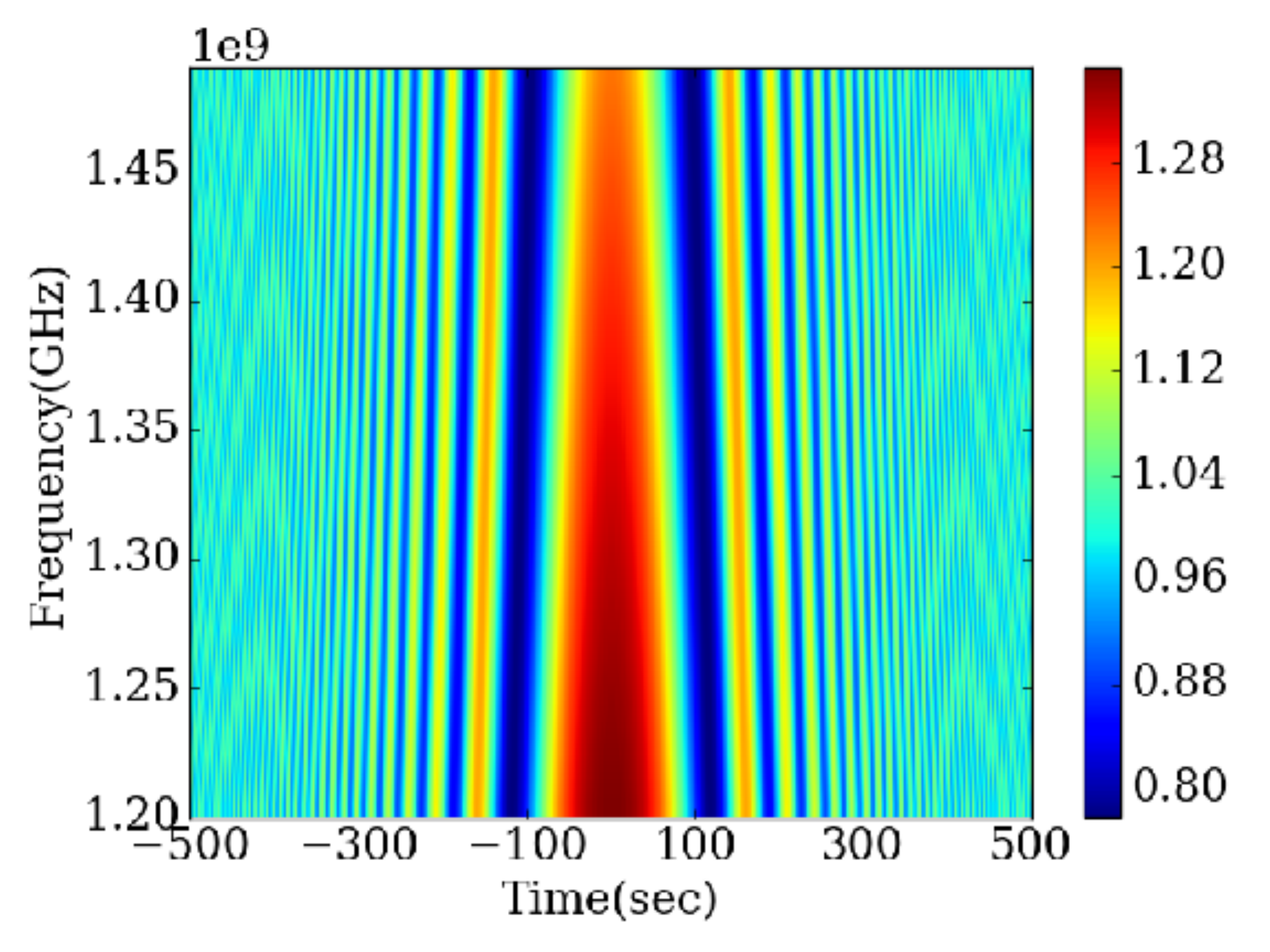}\par
    \includegraphics[width=1.0\linewidth]{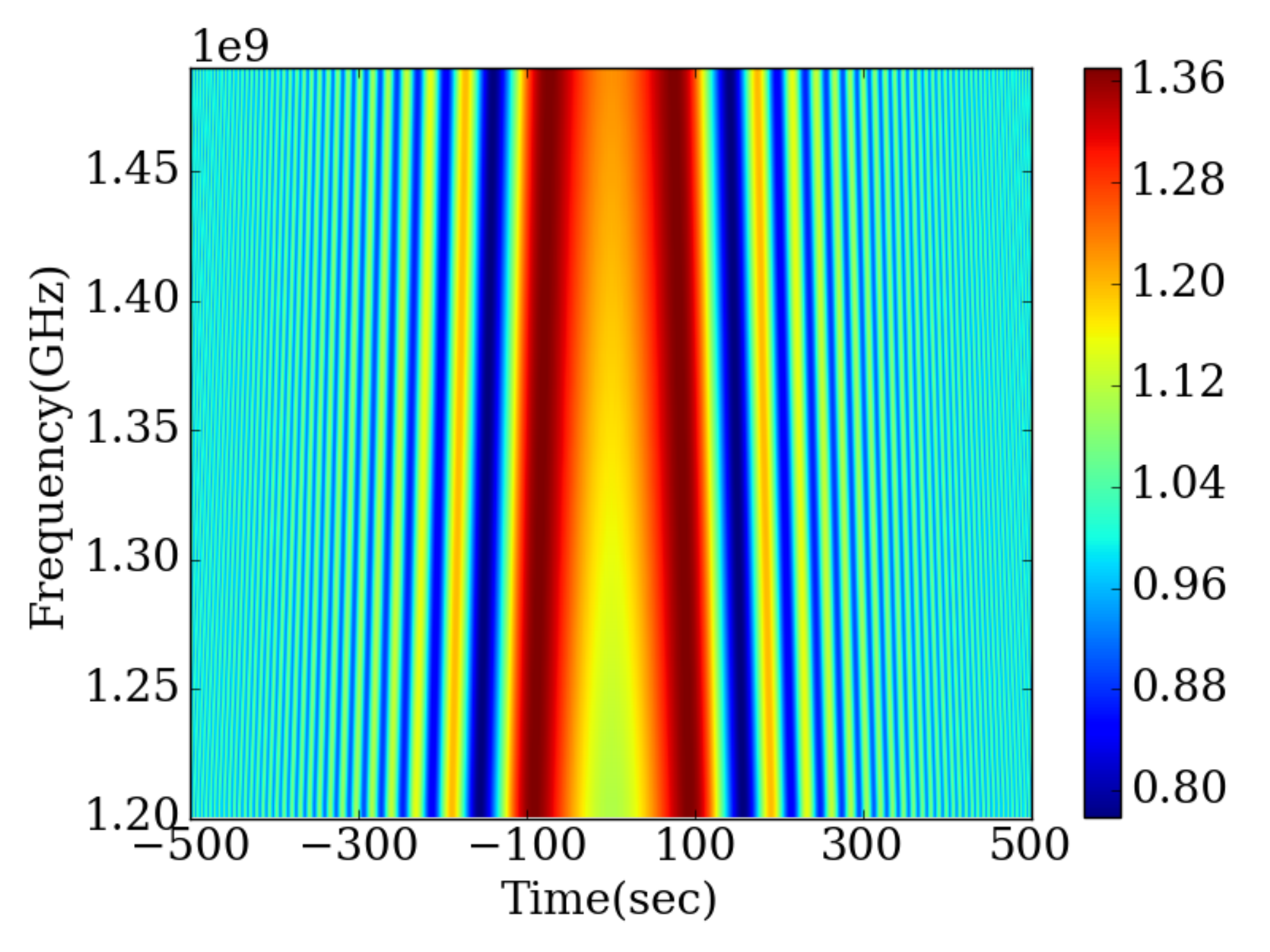}\par
\end{multicols}
\caption{Source at different $x_{min}$ distance from the
edge, $x_{min}$ increasing from top to bottom clockwise, with $x_{min}$
 {\bf $\sim$ 0, 1, 3.2 and 4.8 arc-second} respectively.}
\end{figure*}

Here, we investigate a case in which obstruction grazes 
the source at some minimum distance $x_{min}$, as shown in 
Figure 5. With increasing distance $x_{min}$, the maximum 
intensity keeps decreasing, and approaches to a constant intensity. 
Essentially, the effect of obstruction gradually diminishes 
with increase in $x_{min}$. The dependence of distance $x$ on 
time can be found out readily from the geometry as illustrated in Figure 5. 

\begin{equation}
(x+R)^2= (x_{min}+R)^2+(v^2t^2)
\end{equation}

or

\begin{equation}
x^2+2xR-(x_{min}^2+2x_{min}R+v^2t^2)=0
\end{equation}

where $R$ is the radius of lunar disk, 
and $t$ is the time since closest approach.
Solving the above quadratic equation, one of the solutions we get

\begin{equation}
x= -R+\sqrt{R^2+(x_{min}^2+2x_{min}R+v^2t^2)}
\end{equation}

Figure 6 depicts 4 cases of the dynamic spectra, 
as the minimum distance to obstruction is gradually increased. 
It is evident from the plots that, as the minimum distance 
from the edge increases, the fluctuations in intensity decrease, 
and effects of diffraction vanish gradually. Here again, 
the distance and time are related to each other in non-linear way. 
The factors which can affect the power-law dependence on 
frequency are the minimum distance, velocity and the radius 
of aperture/blockage. The dispersion found in this case doesn't 
resemble the dispersion due to ISM, 
for any of the explored combinations of parameters.

%%Use table environment for a table in one column

%%Use table* environment to get the table spanning both the columns

%\begin{table*}[htb]
%\tabularfont
%\caption{Caption text here}\label{secondTable}
%\begin{tabular}{lccccccccccccr}
%\topline
%\textbf{head1}&\multicolumn{11}{c}{\textbf{head2}}&\textbf{head3}\\
%\midline
%one& two &three&four&five&six&seven&eight&nine&ten&eleven&twelve&thirteen\\
%1&2&3&4&5&6&7&8&9&10&11&12&13\\
%aaa&bbbb&cccc&ddddd&eee&ffff&ggggg&hhhhhhhh&iiii&kkkkkk&hhh&jjjjjj&lllll\\
%\hline
%\end{tabular}
%\tablenotes{Table footnote here. Table spanning both the columns.}
%\end{table*}

%%An example of a figure

%\begin{figure}[!t]
%\includegraphics[width=.8\columnwidth]{fig1.eps}
%\caption{caption goes here}\label{figOne}
%\end{figure}

%%An example of a double column figure
%%Use figure* environment

%\begin{figure*}
%\centering\includegraphics[height=.15\textheight]{fig1.eps}
%\caption{caption spanning two columns}
%\centering\includegraphics[height=.25\textheight]{fig1.eps}
%\caption{caption here}
%\end{figure*}

\subsection{Effect of angular size of the source}

So far, we have considered a point source occultation. 
However, the resultant diffraction pattern has multiple fringes. 
To closely relate to a single dispersed feature, we appeal 
to effects of finite source size. 
 Unless mentioned otherwise, we assume
for simplicity that the source angular structure
(brightness distribution) is independent of frequency 
across the band being considered.
When the apparent source size becomes bigger, 
the fringes get increasing washed out and the dependence 
on frequency changes gradually as we keep increasing 
the size of source. The dispersion delay between two points 
also gets reduced. If the source is large enough, then the 
secondary diffraction fringes vanish completely as shown in Figure 7.
This can be probed by using several different values 
of source size and finding the power index for reference 
points having same intensity.

\begin{figure}
\centering
\includegraphics[width=1.0\linewidth]{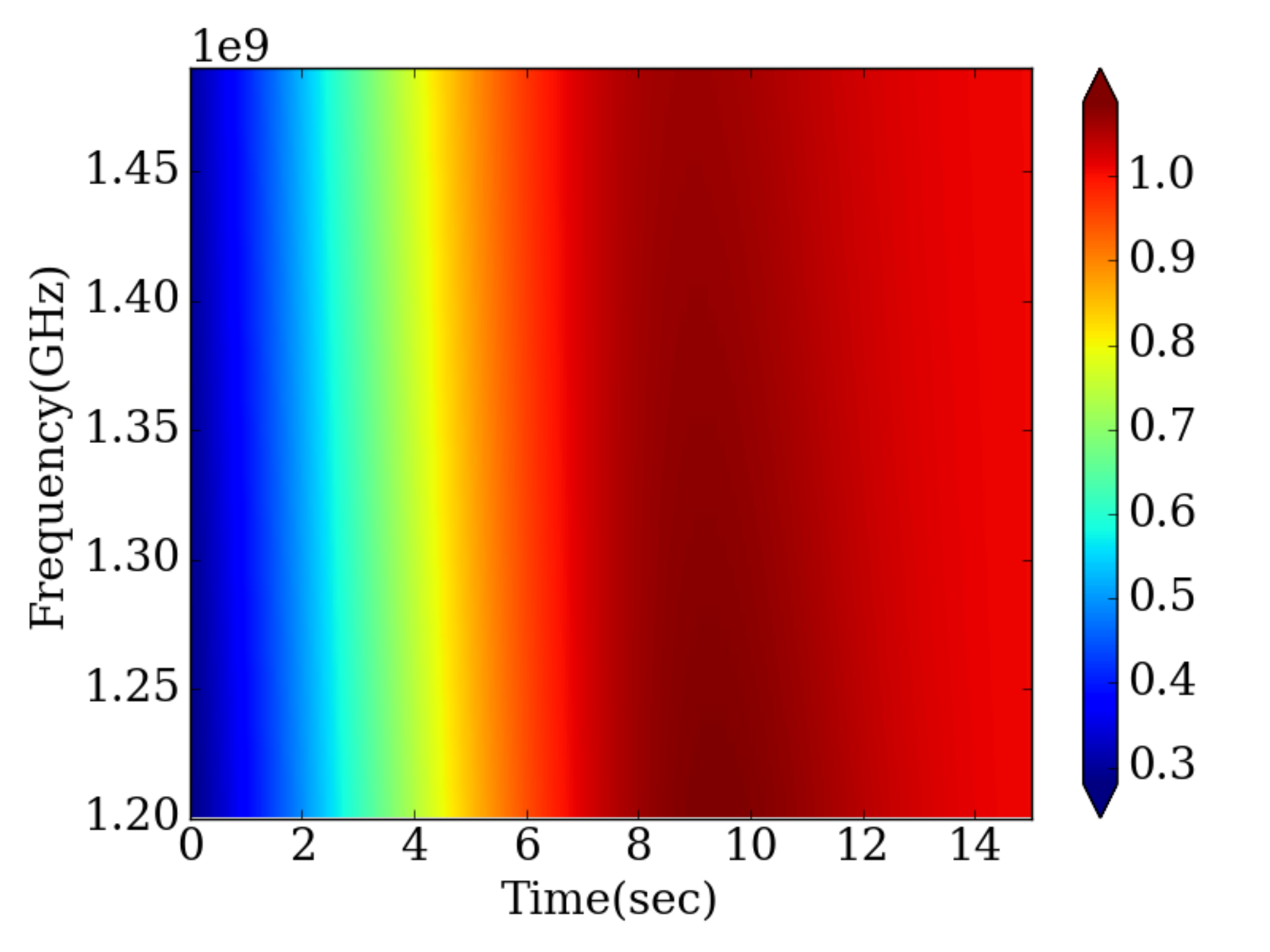}
\caption{Dynamic spectrum for source size of {\bf $\sim$4.2} arc-second, 
wherein the higher order Fresnel lobes or fringes are washed out, as expected.}
\label{fig:test}
\end{figure}

\begin{figure}
\centering
\includegraphics[width=1.0\linewidth]{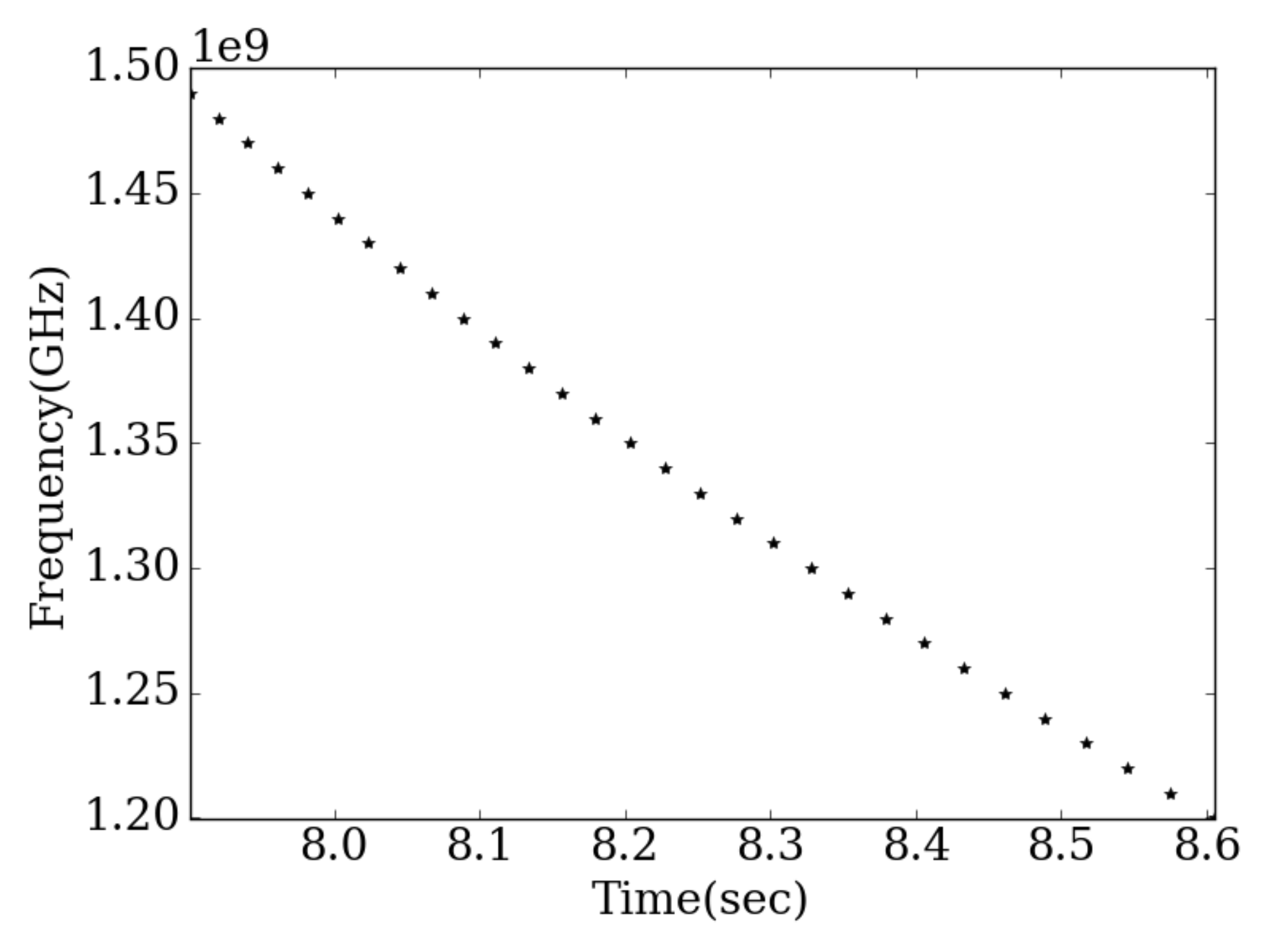}
\includegraphics[width=1.0\linewidth]{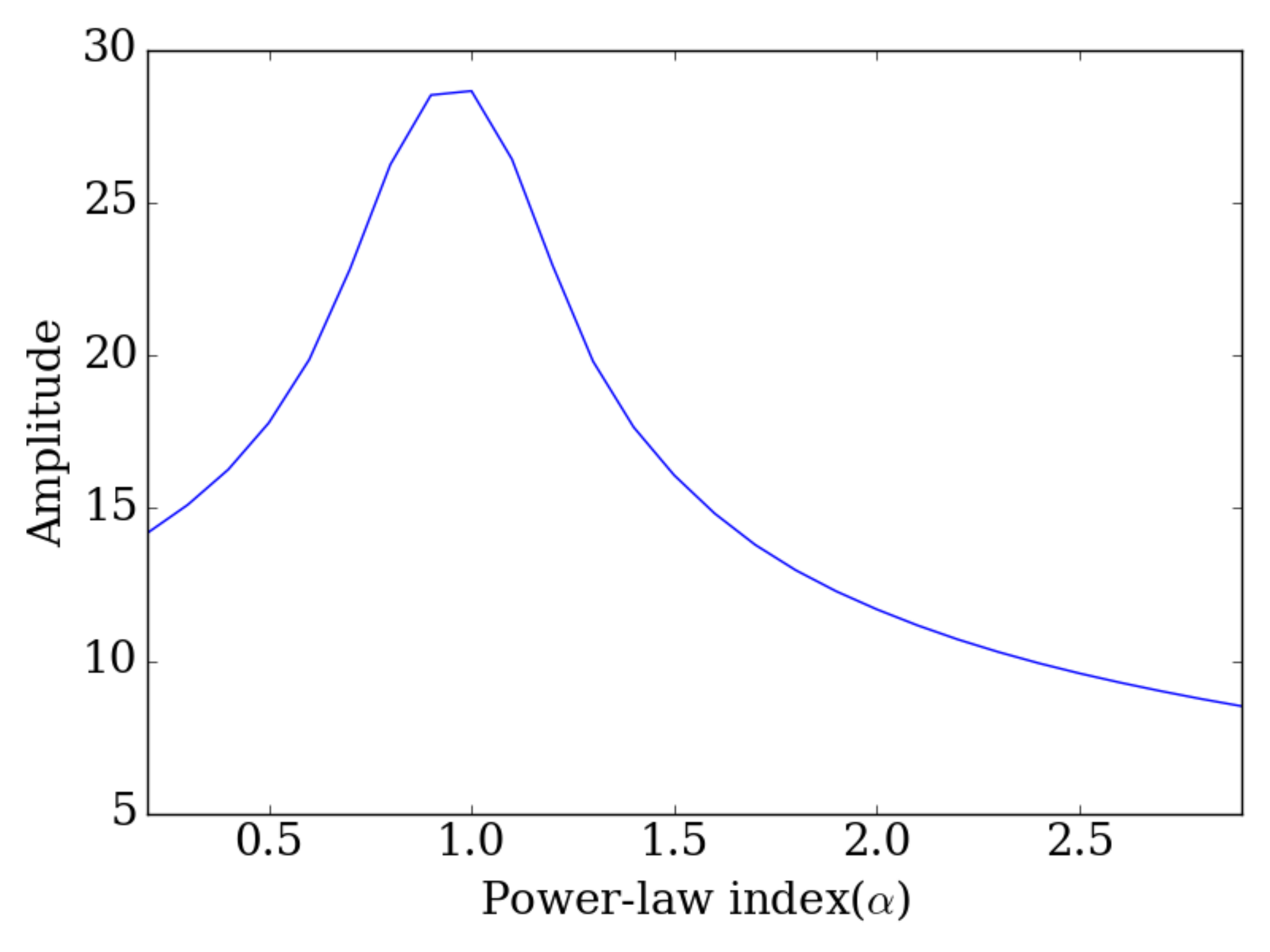}
\includegraphics[width=1.0\linewidth]{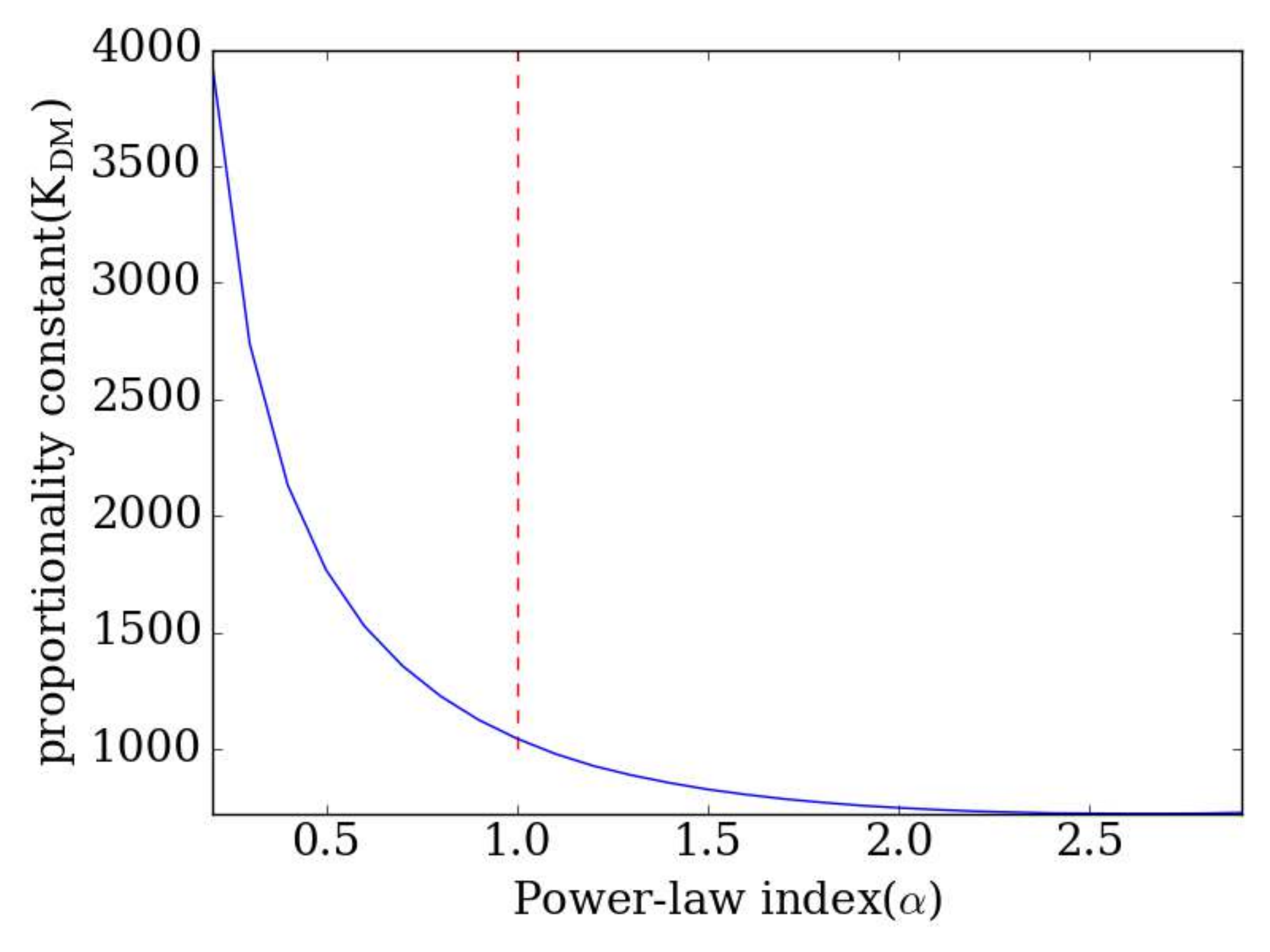}
\caption{The top panel shows the apparent dispersion trend, 
wherein the times corresponding to the peak in intensity 
(in the light curves) shift systematically with frequency. 
The middle and the bottom panels show how the amplitude of 
the dedispersed pulse peak, and the implied DM-like 
proportionality constant ($K_{DM}$), respectively, 
vary as a function of the trial power-law index $\alpha$. 
Here, the assumed source size is {\bf $\sim$3.2} arc-second.}
\label{fig:test}
\end{figure}

\begin{figure}
\centering
\includegraphics[width=1.0\linewidth]{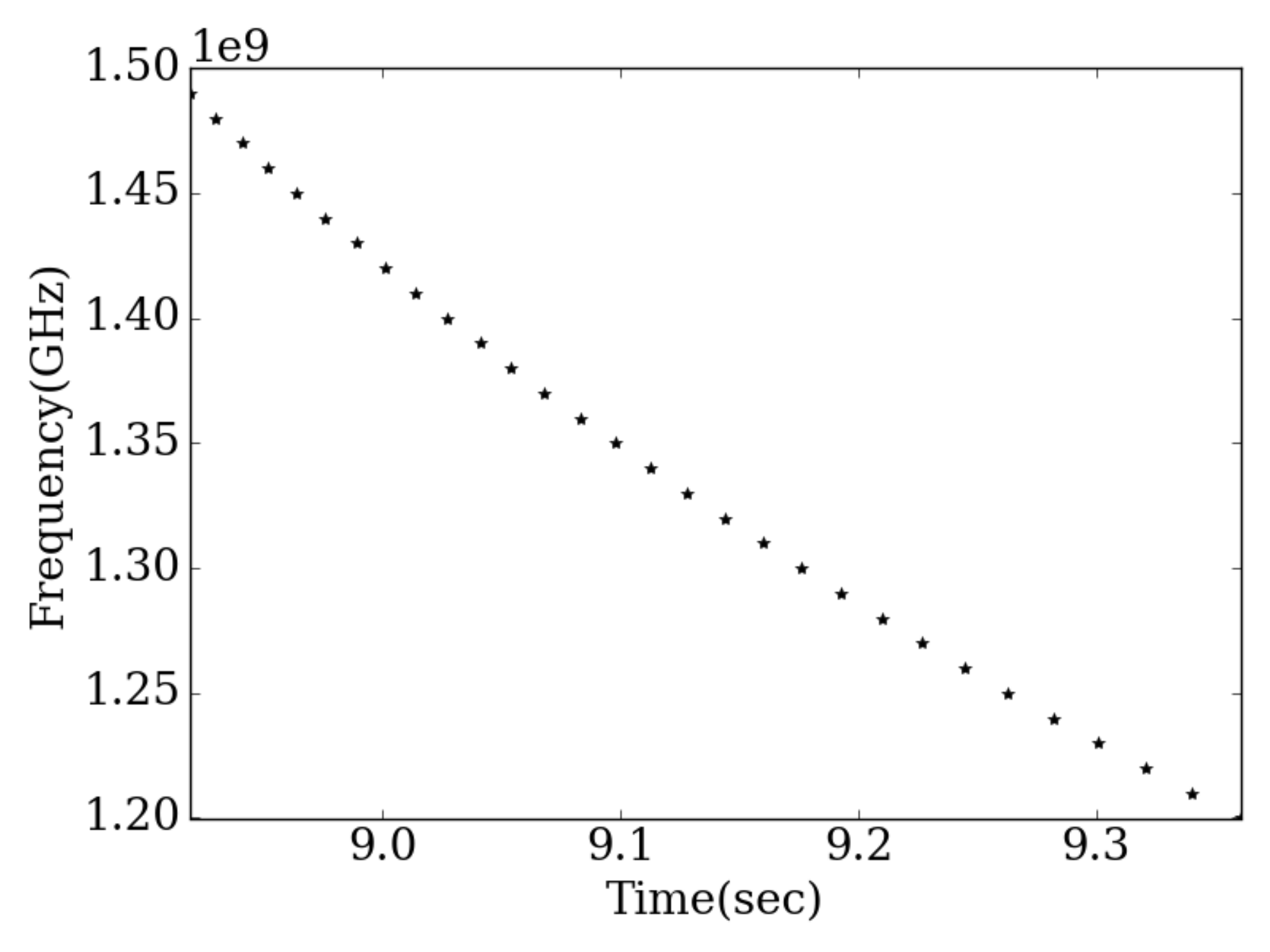}
\includegraphics[width=1.0\linewidth]{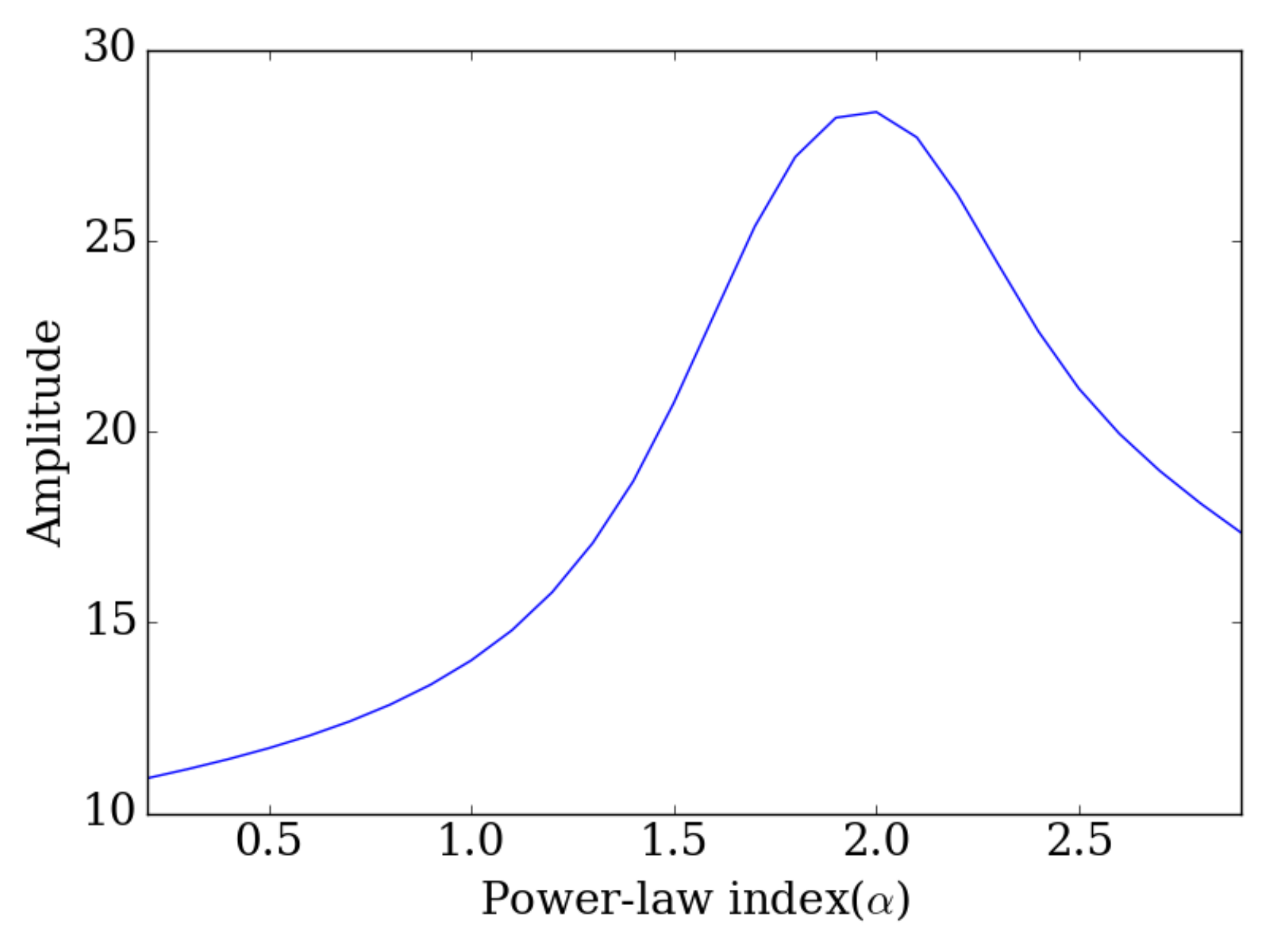}
\includegraphics[width=1.0\linewidth]{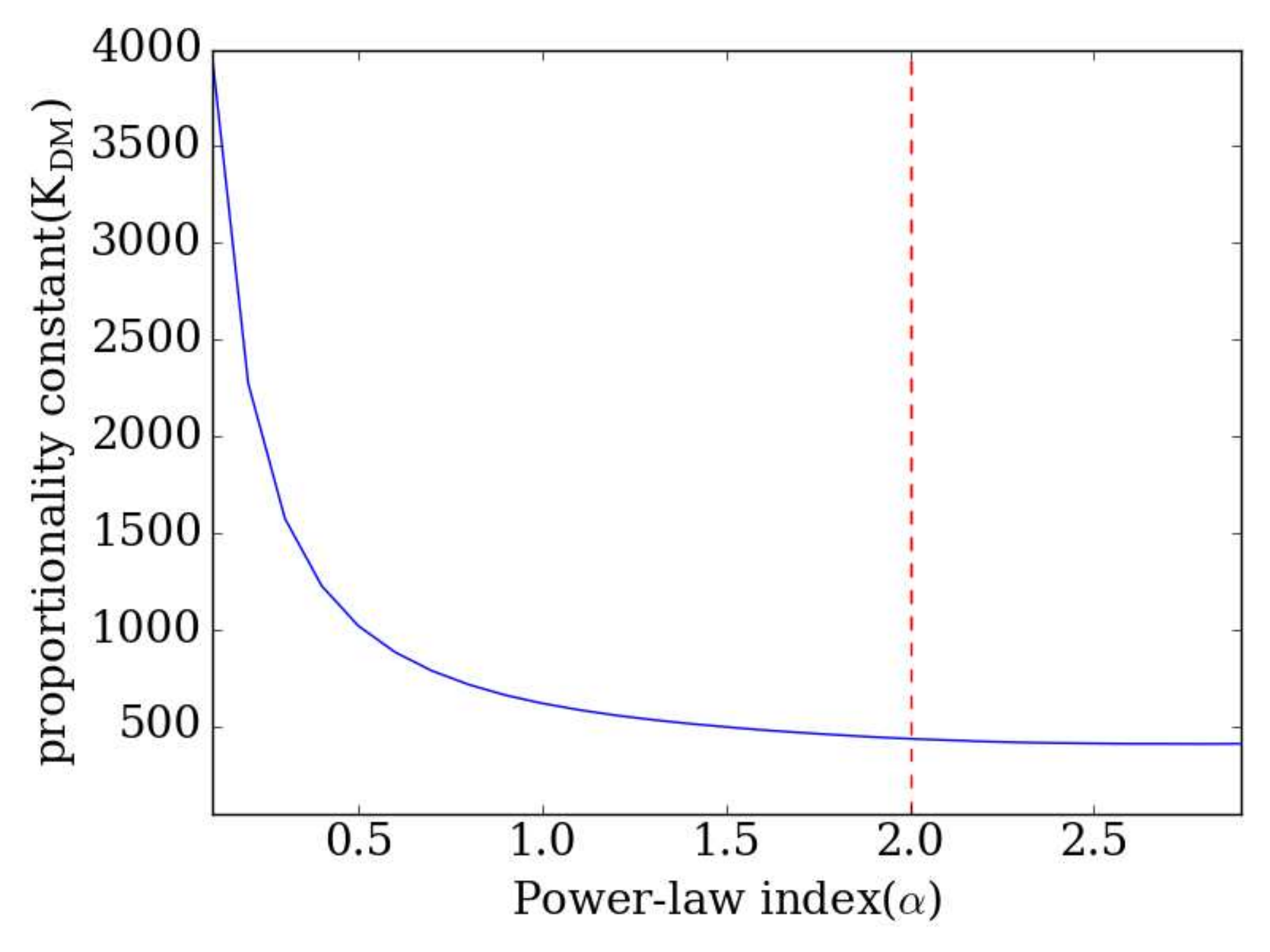}
%\caption{Position of dispersed peak intensity, power-law index and proportionality constant for source size 0.015 arcsecond after dedispersion}
\caption{Similar to Figure 8, but for source size {\bf $\sim$4.2} arc-second.}
\label{fig:test}
\end{figure}

From Figures 8 and 9, we summarize the result of this analysis, 
that as the size of the source varies, the relationship of 
relative time delay with frequency changes. As source grows bigger, 
the fluctuation in intensity diminishes as expected, 
but also does the relative time delay. The apparent net decrease 
in relative time delay, despite the increase in the power-law index, 
is to be understood as due to a more than compensating decrease 
in the DM-like proportionality constant, 
effectively reducing the dispersion-like effect.

\section{Improvement in Signal-to-Noise ratio for large bandwidth}

 The effectiveness of Lunar occultation to probe angular structure of
the occulted sources hinges on high fidelity in measurement of the
diffraction pattern at a chosen wavelength. The high fidelity 
implies the requirement not only of high signal-to-noise 
ratio in the measurements, but also
of preserving the fine details in the apparent temporal pattern, or the
so-called light curve. In radio astronomy measurements 
devoid of fine-scale spatio-temporal 
structure, the signal-to-noise ratio improvement is routinely 
sought by appropriate increase in one or more
of the relevant parameters, such as the aperture area of the telescope, 
pre-detection bandwidth, and post-detection time
constant (or integration) of the receiver. In contrast, the situation
in occultation observations is rich in spatio-temporal structure that
is also explicitly chromatic. This limits the potential
improvements in the contrast against noise, in the occultation observations.
Any increase in the above mentioned parameters beyond their inherent scales
in the occultation signal would wash out the related detail.
These aspects have already been well-appreciated even at early times
(for example, see Scheuer 1965) and optimum choices of
the parameters and estimation method have been made where possible
(see Singal 1987, and references therein). 
In the following, we focus only on the chromatic nature of the
occultation signature, and its implication for broad-band measurements.
And as before, we assume, for simplicity, that the source structure (including
its intensity) does not vary significantly across the observing band. 

In the Lunar occultation observations in the past, 
the resultant diffraction patterns 
observed and used for further analysis have invariably been the
band-averaged versions, as far as we could find. 
As discussed also in the earlier part of this paper,
the chromatic nature of the diffraction signature relating to occultation 
implies a well-defined spatial, and consequent temporal, scaling as
a function of frequency.
Hence, if light curves for a range of frequencies within an observing band
are averaged without accounting for the chromatic effect,
significant undesired smearing of the desired details in the light curve is
inevitable. 
These details and their contrast (i.e. signal-to-noise ratio)
are critical to deciphering the underlying source structure. 
While finite size/structure of sources also causes smearing in the
apparent light curve, it is important to appreciate that here the
smearing function is independent of the distance from the obstruction. 
To illustrate the effect of averaging
over a bandwidth without correcting scaling effect, 
let us consider a point source and 
a finite size source observed at different bandwidths. 
The left panels in Figure 10 \& 11 depict dynamic 
spectra of averaged diffraction pattern over different 
bandwidths for point source and for a source with finite diameter. 
It can be observed that there is more smearing of 
fringes when averaged over larger bandwidths. 
This smearing results in decrease in Signal-to-Noise Ratio (SNR) 
and also, some of the details in the pattern are washed out,
compromising the effective resolution and ability to decode the source structure.

\begin{figure*}[h]
\begin{multicols}{2}
    \includegraphics[width=1.0\linewidth]{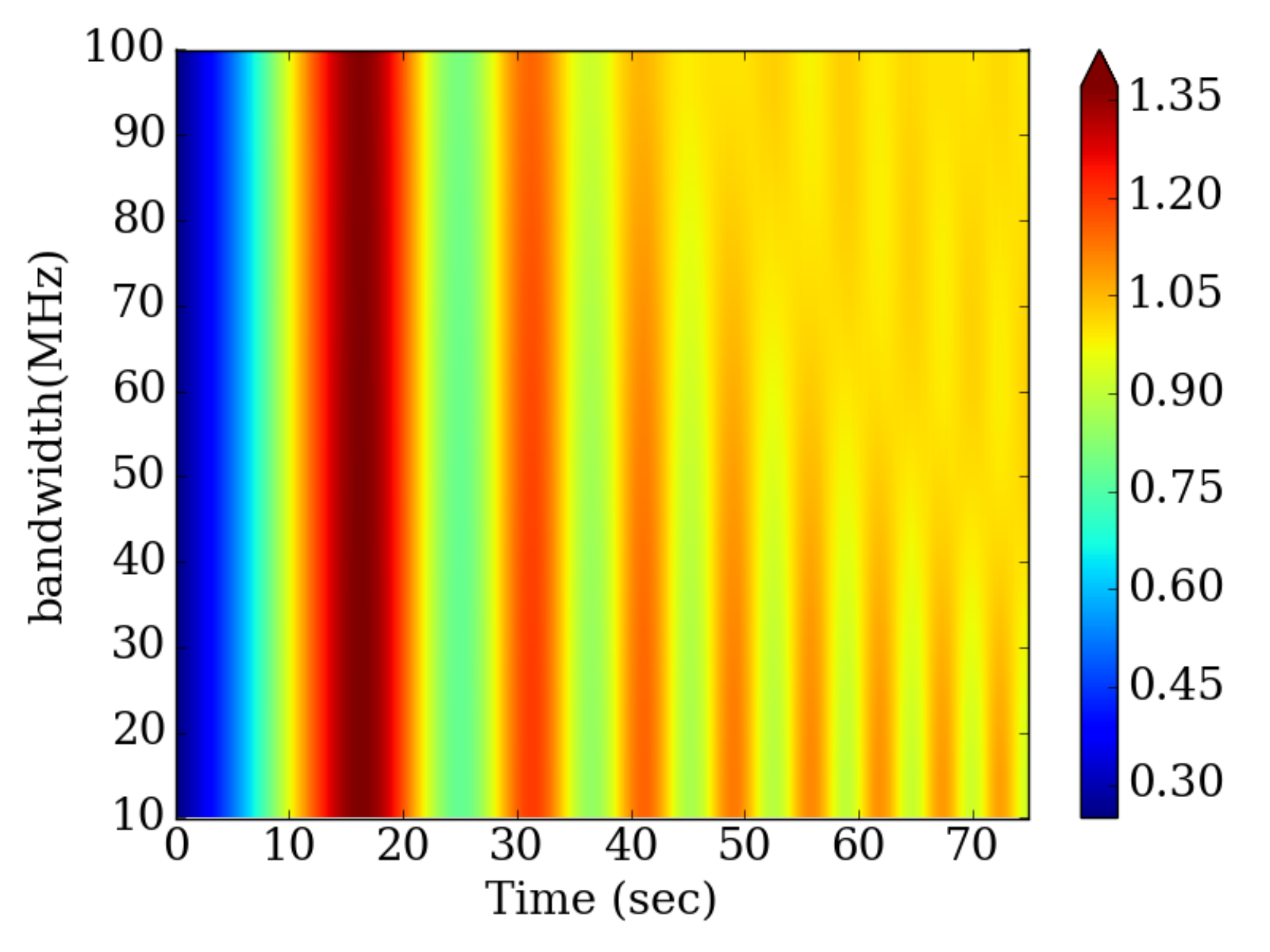}\par 
    \includegraphics[width=1.0\linewidth]{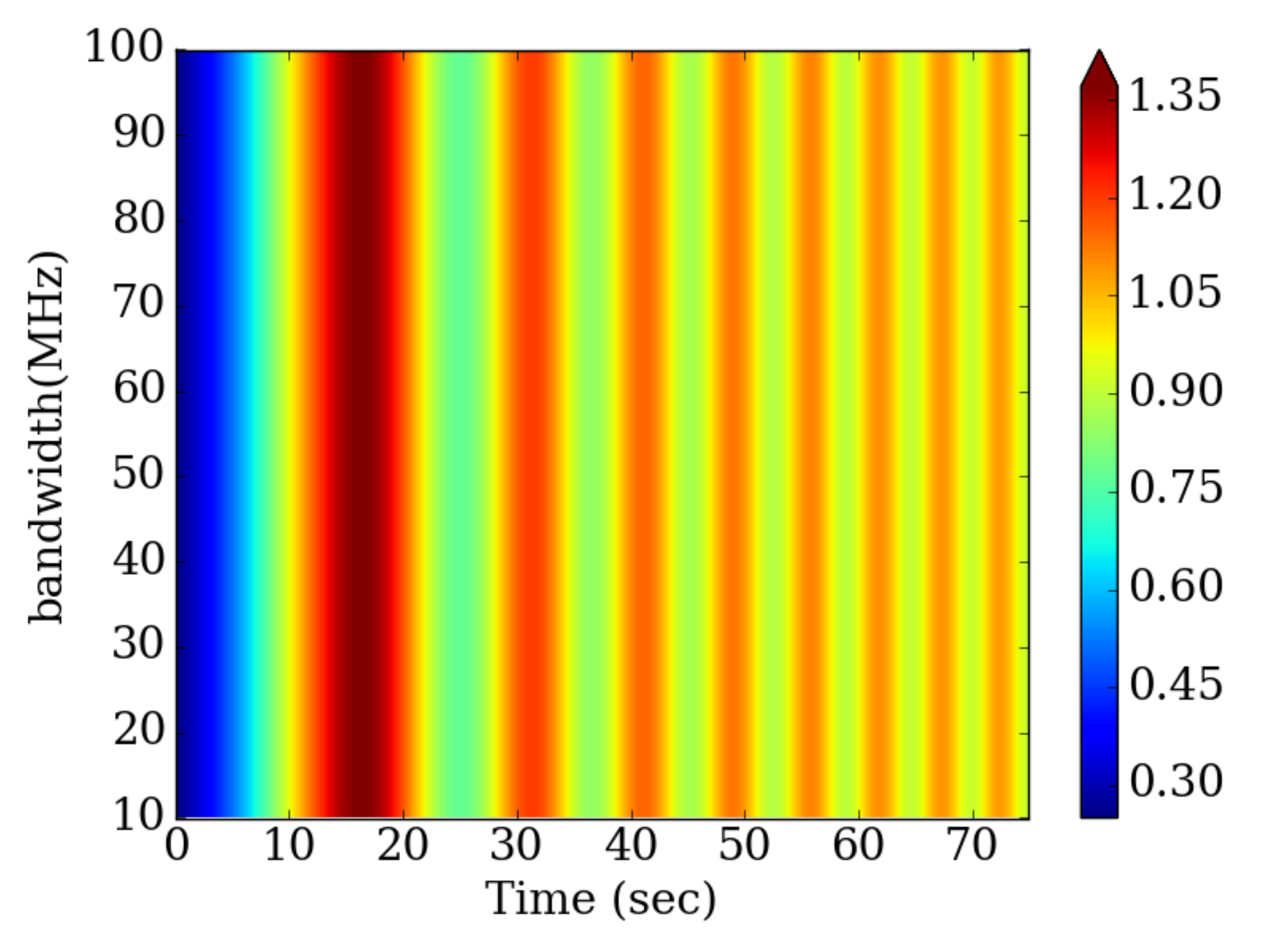}\par 
    \end{multicols}
    
\caption{Above 2-D plot panels depict the band-averaged versions of
Fresnel diffraction curve as a function of time and bandwidth. 
The left \& the right panels represent band-averaging of diffraction 
curves for point source  
without and with frequency scaling correction, respectively.
Despite the changing bandwidth, the center frequency of the bands
is assumed to be fixed at 326 MHz.}    
    
\begin{multicols}{2}
    \includegraphics[width=1.0\linewidth]{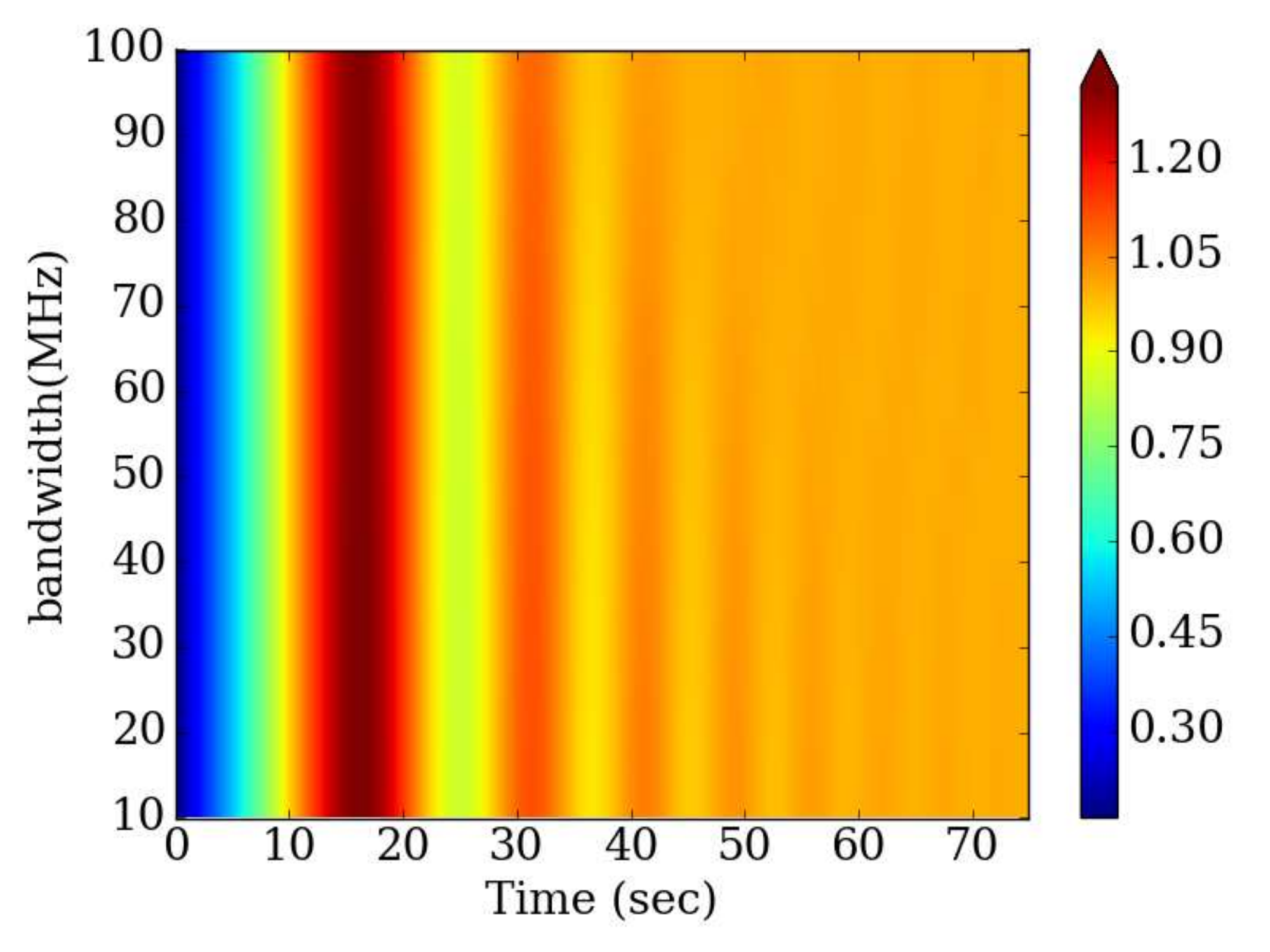}\par
    \includegraphics[width=1.0\linewidth]{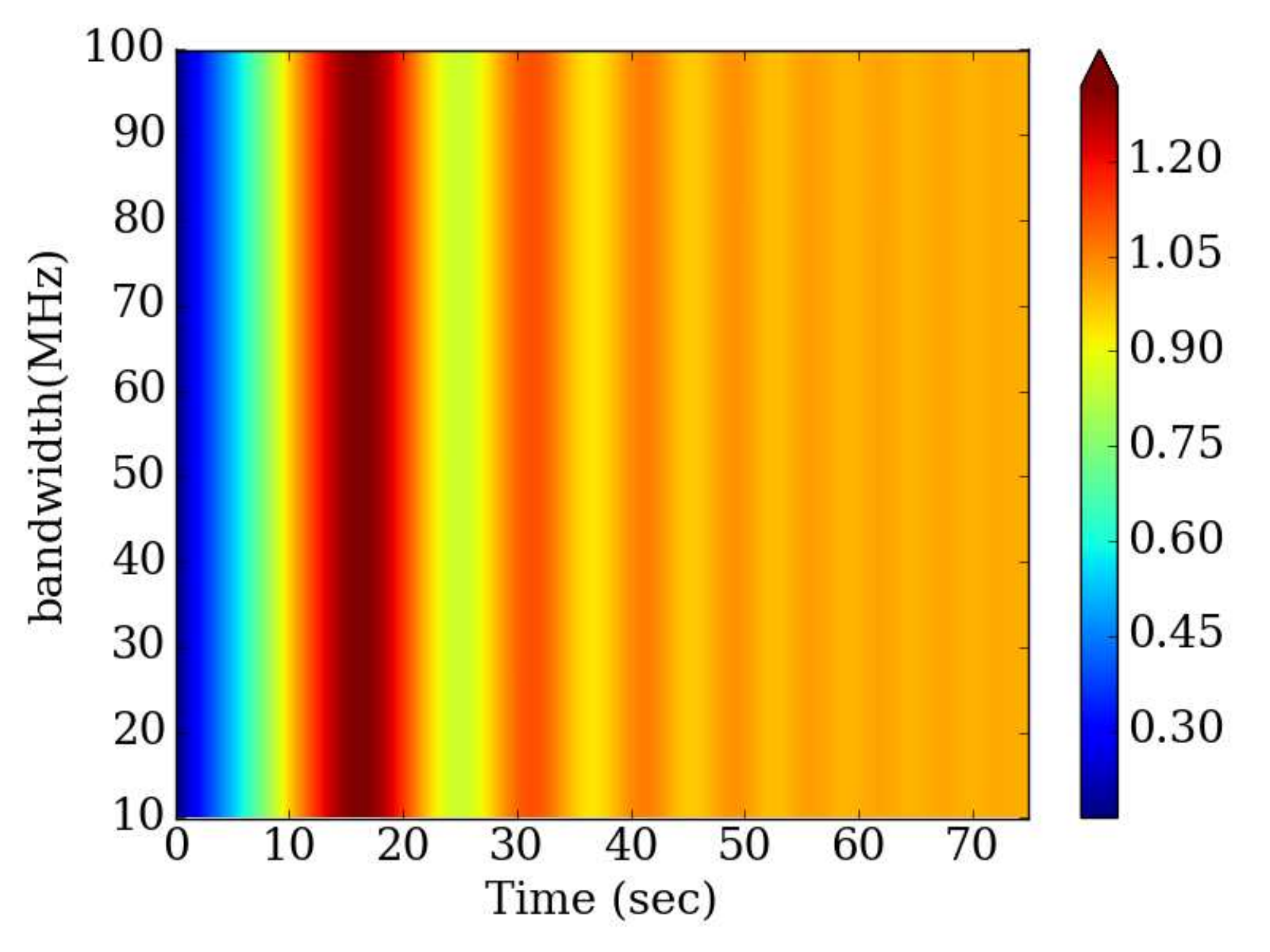}\par
\end{multicols}
\caption{Similar to Figure 10, but for a source size of about 2.5 arc-second (FWHM)}
\end{figure*}

In order to improve the SNR for wide bandwidths, while retaining the {\it fringe} contrast,
suitable scaling correction for diffraction patterns at different 
frequencies can be made before directly averaging them across bandwidth. 
The scaling in time for point size sources follow the relation 
given in Equation 3. The equation can be used for correcting 
scaling effect by taking the scaling with respect to the center frequency 
as reference. After temporal-scaling correction, when diffraction patterns 
are averaged over bandwidth, the underlying signal contrast can be restored. 
The right-side panels in Figure 10 \& 11 show spectrogram 
after such scaling correction. For finite size sources, it might appear that
the averaging without scaling correction suffers 
relatively less deterioration in the intensity pattern than that for a point source.
This is because some smearing of the pattern is already caused by the finite source size. 
However, the scaling correction is required even in this case, for reducing
the effect of {\it decorrelation} due to large bandwidth, which otherwise
would be interpreted wrongly as due to source structure. 
The significant recovery in the fringe contrast as a result of the suggested
correction is reassuring, and recommended. 
Alternatively, it is possible to consider appropriate model fitting path to
decipher source structure from the observed dynamic spectra directly, since
the dynamic spectral signature for a point source can be computed in detail readily.
In the latter case, one can also relax the assumptions about the source structure
being constant across the observed band.
Thus, in any case, more sensitive probes
using lunar occultation appear possible with use of wide bandwidths, offering
sensitivity improvement proportional to square-root of the bandwidth, without
degradation due to decorrelation of the patterns across the band.
Such approach is now routinely feasible with the recent advance in sampling and analysis
of wide-band signals to obtain dynamic spectral data with desired resolutions in
time and frequency.  
 
\section{Conclusion and discussion}
The study of lunar occultation across a broad band 
is conducted, computing and analysis, the
dynamic spectra for the intensity pattern, 
as function of frequency and time. The intensity patterns
show dispersion clearly, which increases as we move away from the edge
of obstruction.The following five cases have been discussed.

(1) The velocity of obstruction is constant.
 
(2) Relative velocity is not constant 

(3) Grazing occultation.  

(4) Apparent source size is finite.

(5) SNR considerations while using large bandwidth, and possible improvements.

The dispersion trend apparent in case of diffraction 
is very different from
ISM dispersion law.
Instead, the diffraction pattern shows spatial 
dispersion delay proportional
to square root of frequency. While considering the first case, 
the dispersion characteristics of Fresnel diffraction 
remain unchanged due to uniform velocity of obstruction. 
By introducing the non-linearity in relation between 
time and distance through non-uniform motion of obstruction, 
ISM like dispersion trend is achievable. 
But the considered scenario may not be physically realizable. 
The third case considered is the most probable and  
frequent situation, but doesn't display dispersion trend 
like ISM at all. In all above cases the occultation of 
point source has been considered. If the source has 
considerable angular width, the secondary fringes can 
be washed out and the dispersion trend closely follows 
ISM dispersion law. Although ISM like dispersion law 
is achievable under few circumstances and single pulse tracks 
are possible by considering effects of suitable 
angular size of the occulted source, but the narrowness of 
FRB pulses is not achievable in case of occultation. 
Also for every positive dispersion signature a negative 
counterpart preceding is unavoidable. 
 Finally, as an independent discussion,
the effect of bandwidth on SNR is considered, noting 
that diffraction pattern would be smeared when 
averaged across large bandwidth. The suggested way of doing 
scaling correction shows how the finer details lost due to bandwidth
decorrelation can be recovered, while improving the
SNR due to increase in bandwidth.

%%Appendix

%%use \balance somewhere in the left column of the last page to balance the two columns in the end page

%%References section


\begin{thebibliography}{99}  
\bibitem{latexcompanion} 
Von Hoerner, S. 1963, {\it ApJ}, {\bf 140}, 65.
%S. Von Hoerner, “LUNAR OCCULTATIONS OF RADIO SOURCES”
%Astrophysical Journal, vol. 140, p. 65, December, 1963.
\bibitem{latexcompanion} 
Hazard, C. 1976, “Lunar Occultation Measurements”, Elsevier (1976), vol 12(c),
pp. 92-117.
%C. Hazard, “Lunar Occultation Measurements”, Elsevier, vol 12, part c,
%pp. 92-117, 1976.
\bibitem{latexcompanion} 
Lorimer, D. R., et al. 2007, {\it Science}, {\bf 318 (5851)}, 777.
%D. R. Lorimer, et al , "A Bright Millisecond Radio Burst of Extragalactic Origin", Science, Vol. 318, Issue 5851, pp. 777-780, 2007.
\bibitem{latexcompanion} 
 Swarup, G, et al. 1971, {\it Nature Physical Science}, {\bf 230}, 185 
\bibitem{latexcompanion} 
 Scheuer, P. A. G., 1962, {\it Australian Journal of Physics}, {\bf 15}, 333 
\bibitem{latexcompanion} 
 Scheuer, P. A. G., 1965, {\it Mon. Not. R. Astr. Soc.}, {\bf 129}, 199 
\bibitem{latexcompanion} 
Ghatak, A. 2010, "OPTICS", McGraw-Hill (2010), pp. 314-317.
%A. Ghatak, "OPTICS", McGraw-Hill, 2010, pp. 314-317
\bibitem{latexcompanion} 
 Hazard, C., et al. 1963, {\it Nat}, {\bf 197 (4872)},pp.1037-1039.
\bibitem{latexcompanion} 
Singal, A. K., 1987, {\it A\&AS}, {\bf 69},pp. 91-115 
      

\end{thebibliography}
\end{document}